%% file: main.tex
\documentclass[10pt,journal,compsoc]{IEEEtran}
\usepackage{amsmath,amsfonts}
\usepackage{algorithmic}
\usepackage{algorithm}
\usepackage{array}
\usepackage[caption=false,font=normalsize,labelfont=sf,textfont=sf]{subfig}
\usepackage{textcomp}
\usepackage{stfloats}
\usepackage{url}
\usepackage{verbatim}
\usepackage{graphicx}
\usepackage{hyperref}
\usepackage{xcolor}
\usepackage{amsthm}
\allowdisplaybreaks[4] 

\ifCLASSOPTIONcompsoc
  \usepackage[nocompress]{cite}
\else
  \usepackage{cite}
\fi
\ifCLASSINFOpdf
\else
\fi
\begin{document}
%
\title{Blind Image Deconvolution Using Variational Deep Image Prior}
%
%
%
%

\author{Dong Huo, 
        Abbas Masoumzadeh, 
        Rafsanjany Kushol, 
        and Yee-Hong Yang,~\IEEEmembership{Senior Member,~IEEE}
\IEEEcompsocitemizethanks{\IEEEcompsocthanksitem D. Huo, A. Masoumzadeh, R. Kushol and Y. Yang are with the Department of Computing Science, University of Alberta, Edmonton, AB T6G 2R3, Canada (e-mail:
dhuo@ualberta.ca; a.masoumzadeh@ualberta.ca; kushol@ualberta.ca; herberty@ualberta.ca)\protect\\
}
}

%
%

\markboth{Journal of \LaTeX\ Class Files,~Vol.~14, No.~8, August~2015}%
{Shell \MakeLowercase{\textit{et al.}}: Bare Demo of IEEEtran.cls for Computer Society Journals}
%



\input{0_abstract}

\maketitle

\IEEEdisplaynontitleabstractindextext

%
\IEEEpeerreviewmaketitle

\input{1_introduction}
\input{2_review}
\input{3_method}
\input{5_experiments}
\input{6_conclusion}
\input{7_acknowledgment}

\ifCLASSOPTIONcaptionsoff
  \newpage
\fi



%

\bibliographystyle{IEEEtran}
\bibliography{bib}

\clearpage
\onecolumn
\appendix
\input{4_derivation}

\end{document}

%% file: 0_abstract.tex
\IEEEtitleabstractindextext{%
\begin{abstract}
Conventional deconvolution methods utilize hand-crafted image priors to constrain the optimization. While deep-learning-based methods have simplified the optimization by end-to-end training, they fail to generalize well to blurs unseen in the training dataset. Thus, training image-specific models is important for higher generalization. Deep image prior (DIP) provides an approach to optimize the weights of a randomly initialized network with a single degraded image by maximum a posteriori (MAP), which shows that the architecture of a network can serve as the hand-crafted image prior. Unlike conventional hand-crafted image priors, which are obtained through statistical methods, finding a suitable network architecture is challenging due to the unclear relationship between images and their corresponding architectures. As a result, the network architecture cannot provide enough constraint for the latent sharp image. This paper proposes a new variational deep image prior (VDIP) for blind image deconvolution, which exploits additive hand-crafted image priors on latent sharp images and approximates a distribution for each pixel to avoid suboptimal solutions. Our mathematical analysis shows that the proposed method can better constrain the optimization. The experimental results further demonstrate that the generated images have better quality than that of the original DIP on benchmark datasets. The source code of our VDIP is available at \href{https://github.com/Dong-Huo/VDIP-Deconvolution}{https://github.com/Dong-Huo/VDIP-Deconvolution}. 
\end{abstract}

}

%% file: 1_introduction.tex
\IEEEraisesectionheading{\section{Introduction}\label{sec:introduction}}
\IEEEPARstart{B}{lind} image deconvolution is aimed at recovering the latent sharp image based on a single blurred image without knowing the blur kernel. When the blur kernel is spatially invariant, it can be modeled as 
\begin{equation}
	I_b= k \otimes I_s + n,
	\label{eqn:convolution_model}
\end{equation}   
where $I_b$ denotes the blurred image, $k$ the blur kernel, $\otimes$  the convolution operator, $I_s$ the latent sharp image and $n$ the additive noise. Most conventional methods utilize maximum a posteriori (MAP) to alternatively solve for $k$ and $ I_s$, which is formulated as 
\begin{equation}
\underset {I_s, k} {\text{arg max}} P(I_s, k | I_b) = \underset {I_s, k} {\text{arg max}} P(I_b | I_s, k)P(I_s)P(k)
\label{eqn:map}
\end{equation}   
where $P(I_b | I_s, k)$ is the likelihood term, $P(I_s)$ and $P(k)$ are the prior distributions of the latent sharp image and the blur kernel, respectively.

Conventional methods propose various priors to solve the problem~
\cite{joshi2008psf, shan2008high, chan1998total, perrone2015clearer, vsroubek2020motion, babacan2012bayesian, krishnan2009fast, joshi2009image, xu2013unnatural, levin2009understanding, michaeli2014blind, pan2016blind, yan2017image, yang2021blind, dong2017blind, cho2009fast, chen2019blind, bai2019single}. 
Among them, the sparse image prior is one of the most widely used priors in image deconvolution, which includes special cases such as the Gaussian prior~\cite{cho2009fast}, the total variational (TV) prior~\cite{chan1998total}, and the hyper-Laplacian prior~\cite{krishnan2009fast}. Fergus~\textit{et al.}~\cite{fergus2006removing} illustrate experimentally that the sparse image prior with MAP (sparse MAP) often removes almost all of the gradients. Levin~\textit{et al.}~\cite{levin2009understanding} also demonstrate that the sparse MAP is more likely to generate the original blurred image than the latent sharp image when normalizing the blur kernel. In other words, the estimated kernel is more likely to be a delta kernel. Even when the estimated kernel is not a delta kernel, the method is easy to be trapped at a local minimum and hard to escape. Delayed normalization~\cite{perrone2015clearer} can avoid the delta kernel but still suffers from getting trapped at a local minimum.  Edge reweighting~\cite{xu2010two} and edge-selection~\cite{cho2009fast}, which need carefully chosen hyper-parameters, are utilized to address these problems by removing small edges and noise before estimating the kernel. Variational Bayesian (VB) based methods~\cite{fergus2006removing, babacan2012bayesian} remit the issues of the sparse MAP by considering the standard deviation of images. 

Recently, deep-learning-based methods~\cite{nah2017deep, nimisha2017blur, li2018learning, zhang2018dynamic, tao2018scale, kupyn2018deblurgan, zhang2019gated, zhang2019deep, lu2019unsupervised, kupyn2019deblurgan, purohit2020region, suin2020spatially, zamir2021multi, tran2021explore} have been applied to this problem, which can implicitly learn the image prior within the network by training on a large dataset. Due to the high dependency on the training datasets, deep-learning-based methods do not generalize well to some image-specific information~\cite{shocher2018zero} (e.g., blur kernels and features) which is not encountered during training. Thus, it is necessary to learn an image-specific model. 

Deep image prior (DIP)~\cite{ulyanov2018deep} is an appealing approach to optimizing a network using a single degraded image. Indeed, the architecture of a generator network can capture a low-level image prior for image restoration. Ren~\textit{et al.}~\cite{ren2020neural} utilize the DIP to handle  blind image deconvolution, and  formulate the problem as 
\begin{equation}
\begin{aligned}
&\underset {I_s, k, \theta_I, \theta_k} {\text{arg max}} P(I_s, k, \theta_I, \theta_k| I_b)\\
=&\underset {I_s, k, \theta_I, \theta_k} {\text{arg max}} P(I_b | I_s, k)P(I_s | \theta_I)P(k |\theta_k)P(\theta_I)P(\theta_k),
\label{eqn:map_2}
\end{aligned}
\end{equation}   
where $\theta_I$ and $\theta_k$ denote the parameters of the image generator $G_I()$ and of the kernel generator $G_k()$, respectively, $P(\theta_I)$ and $P(\theta_k)$ are, respectively, the priors of these parameters, and $P(I_s | \theta_I)$ and $P(k | \theta_k)$, respectively, are the image and the kernel prior learned by $G_I()$ and $G_k()$. Since they assume that $P(\theta_I)$ and $P(\theta_k)$ are constant, there is no constraint on the generated image and the kernel. As a result, it is not surprising that the outputs are suboptimal. One solution is to apply the sparse image prior to constrain $P(I_s | \theta_I)$, but the method still suffers from the  problems of the sparse MAP similar to that of conventional methods.  

To solve the above mentioned issues of the DIP, we attempt to adopt VB-based methods to the DIP, so that not only the optimization is constrained but also the problems of the sparse MAP can be avoided. Conventional VB-based methods~\cite{fergus2006removing, babacan2012bayesian, yang2019variational} utilize a trivial (e.g., Gaussian) distribution to directly approximate the posterior distribution (the left term of Eqn.~\ref{eqn:map} and Eqn.~\ref{eqn:map_2}) by minimizing the Kullback–Leibler (KL) divergence~\cite{kullback1997information} instead of using MAP. Although the accurate posterior distribution is hard to obtain, the approximated one is good enough and much more robust than the result of MAP. In order to combine the DIP with VB, we propose a new variational deep image prior (VDIP) to learn the distributions of all latent variables (sharp images and blur kernels) which is motivated by the idea of variational auto-encoder~\cite{kingma2013auto}. More details of the mathematical analysis of why VDIP can perform better than DIP are given in Section~\ref{sec:method}.

Our contributions are summarized as follows:
\begin{itemize}
\item{We propose a novel variational deep image prior (VDIP) for single image blind deconvolution by integrating the deep image prior and variational Bayes.}
\item{We provide a complete derivation of our final loss function and a mathematical analysis to demonstrate that the proposed method can better constrain the optimization than that of DIP.}
\item{Our experiments show that the proposed VDIP can significantly improve over the DIP in both quantitative results on benchmark datasets and the quality of the generated sharp images.}
\end{itemize}

%% file: 2_review.tex
\section{Related Work}
\subsection{Blind Image Deconvolution}
Some conventional single image blind deconvolution methods focus on the distribution of image gradient for sparse high-frequency information. Fergus~\textit{et al.}~\cite{fergus2006removing} propose a heavy-tailed natural image prior, which is approximated by a mixture-of-Gaussian model. Shan~\textit{et al.}~\cite{shan2008high} demonstrate that the ringing effect on the deblurred image results from the estimation error of the blur kernel and noise. Cho and Lee~\cite{cho2009fast} utilize the bilateral filter and the shock filter to remove noise and to enhance edges. Xu and Jia~\cite{xu2010two} find that edges smaller than the kernel size are harmful to kernel estimation and propose an r-map to measure the usefulness of edges. Krishnan \textit{et al.}~\cite{krishnan2011blind} adopt the ratio of the L1 norm and the L2 norm to avoid the scale variant prior, which is much closer to the L0 norm. Levin \textit{et al.}~\cite{levin2009understanding} prove that MAP with the sparse image prior favors a blurred solution so that they approximate the marginalization of the blur kernel, which has a closed-form solution when using the Gaussian image prior. Babacan~\textit{et al.}~\cite{babacan2012bayesian} exploit the concave conjugate of the super-Gaussian prior and directly estimate the posterior distribution using VB to avoid the issues of the sparse MAP. Dong~\textit{et al.}~\cite{dong2017blind} adopt a piecewise function to mimic the L0 norm around zero and to smooth out significant outliers, which is similar to the work of Xu~\textit{et al.}~\cite{xu2013unnatural}. Chen~\textit{et al.}~\cite{chen2020enhanced} who enhance the sparse prior by combining the L0 and L1 norm. Yang~\textit{et al.}~\cite{yang2019variational} introduce a restarting technique to further improve the performance of VB-based methods.

Some other conventional methods utilize  properties of images to form priors. Michaeli and Irani~\cite{michaeli2014blind} find that blur significantly decreases cross-scale patch recurrence. Thus, they constrain the output by minimizing the dissimilarity between nearest-neighbor patches. Lai~\textit{et al.}~\cite{lai2015blur} assume that each local patch contains two primary colors, and the distance between them should be maximized by deconvolution. Pan~\textit{et al.}~\cite{pan2016blind} apply the dark channel prior to handle blind deconvolution and achieve good results. Yan~\textit{et al.}~\cite{yan2017image} combine the bright and the dark channel priors to overcome the limitation on bright dominant images. Ren~\textit{et al.}~\cite{ren2016image} derive an enhanced low-rank prior to reduce the number of non-zero singular values of the image. Pan~\textit{et al.}~\cite{pan2019phase} exploit the phase-only image of a blurred image to estimate the start and end point of the blur kernel, which is efficient for linear motion. Bai~\textit{et al.}~\cite{bai2019single} utilize the downsampled blurred image as the prior and recover the latent sharp image from coarse to fine. Chen~\textit{et al.}~\cite{chen2019blind} calculate the bright channel of the gradient maps for deblurring images without enough dark and bright pixels.

Deep-learning-based methods are also applied to this problem. Chakrabarti~\cite{chakrabarti2016neural} trains a network to estimate the Fourier coefficients of blur kernels. Liu~\textit{et al.}~\cite{liu2016learning} and Zhang~\textit{et al.}~\cite{zhang2018dynamic} exploit recursive filters to take advantage of context information. Generative adversarial networks (GANs) are also exploited for faster convergence and better visual quality~\cite{nimisha2017blur, kupyn2018deblurgan, kupyn2019deblurgan}. Gong~\textit{et al.}~\cite{gong2017motion} adopt a network to learn the motion flow. Xu~\textit{et al.}~\cite{xu2017motion} develop a network to generate sharp gradient maps for kernel estimation. To enhance the network output, some utilize  multi-stage strategies, e.g., multi-scale~\cite{nah2017deep, tao2018scale, liu2018recurrent}, multi-patch~\cite{zhang2019deep, suin2020spatially, zamir2021multi} and multi-temporal~\cite{park2020multi}. Asim~\textit{et al.}~\cite{asim2020blind} adopt a well-trained sharp image generator to generate the sharp image closest to the blurred one. Tran~\textit{et al.}~\cite{tran2021explore} develop a sharp image auto-encoder and a blur representation learning network, then two well-trained networks are fixed as a deep generative prior~\cite{asim2020blind}. Li~\textit{et al.}~\cite{li2018learning} adopt a well-trained classifier (which can distinguish blurred images and sharp images) as an extra constraint of the MAP framework, and optimize the problem with the half-quadratic splitting method similar to that used in conventional methods. 

Different from~\cite{li2018learning, asim2020blind, tran2021explore, pan2020dgp_pami, zhang2020plug} in which priors need to be trained on external datasets, our proposed method is optimized with only one single blurred input image and the whole framework is optimized by gradient descent instead of conventional optimization-based methods~\cite{li2018learning}. Although Asim~\textit{et al.}~\cite{asim2020blind} also provide a method optimized with a single image, the method degenerates to the DIP~\cite{ren2020neural} with a sparse image prior and learnable inputs, which cannot avoid the problems of the sparse MAP. As well, none of the mentioned deep-learning-based methods consider the standard deviation of the 
image.
\subsection{Deep Image Prior}
Ulyanov~\textit{et al.}~\cite{ulyanov2018deep} introduce the concept of the deep image prior (DIP) that the structure of a randomly-initialized network can be used as an image prior for image restoration tasks. Ren~\textit{et al.}~\cite{ren2020neural} adopt the DIP to implicitly learn the image prior and the kernel prior for blind image deconvolution. Early stopping with carefully chosen time,  added random noise to the input and to the gradient with fixed noise level are applied to avoid the suboptimal solution of DIP~\cite{cheng2019bayesian}. Neural architecture search (NAS) can help to search for these hyper-parameters heuristically~\cite{ho2020neural}, but with the substantial increase in computational cost. Double-DIP~\cite{gandelsman2019double} can handle the image separation problems, e.g., image segmentation, image dehazing, and transparency separation, but does not perform well for blind image deconvolution~\cite{ren2020neural}. Some methods stabilize the optimization by adding extra priors to the loss function~\cite{abu2021image, mataev2019deepred}. However, this technique only works when the degradation kernel is known.

\subsection{Variational Auto-encoder}
Kingma~\textit{et al.}~\cite{kingma2013auto} introduce the concept of variational auto-encoder (VAE) for image generation. The goal is to learn a model that  generates an image $x$ given a sampled latent variable $z$, which can be formulated as $P(x|z) = P(x) P(z|x) / P(z),$ where $P(x)$ is constant. Since obtaining the true distribution of $P(z|x)$ is nontrivial, they utilize a Gaussian distribution $Q(z)$ to approximate  $P(z|x)$ with a network to learn the expectation and the standard deviation. Thus, the target of VAE can be converted to minimizing the KL divergence between $Q(z)$ and $P(z|x)$. Vahdat~\textit{et al.}~\cite{vahdat2020NVAE} further stabilize the training of VAE by partitioning the latent variables into groups. Similar to  image generation, the target of image deconvolution is learning a model to generate a blurred image $I_b$ given a sampled latent sharp image $I_s$ and a blur kernel $k$, and the distributions of $P(I_s|I_b)$ and $P(k|I_b)$ are learned by the network. And predefined hand-crafted $P(I_s)$ and $P(k)$ can help to constrain the optimization.

%% file: 3_method.tex
\section{Proposed Method}\label{sec:method}
In this section, we provide the mathematical analysis of the feasibility of our proposed methods. More derivation details are given in the supplementary materials.
\subsection{Super-Gaussian Distribution}
Conventional image priors can be formulated as a super-Gaussian distribution:
\begin{gather}
P(I_s) = W\exp\left(-\frac{\rho(F_x(I_s)) + \rho(F_y(I_s))}{2}\right),
\label{eqn:prior}
\end{gather}   
where $W$ is the normalization coefficient, and $\rho()$ is the penalty function to constrain the sparsity of $F_x(I_s)$ and $F_y(I_s)$. For sparse image priors, $F_x()$ and $F_y()$ are gradient kernels $[-1, 1]^T$ and $[-1, 1]$. When $\rho()$ is quadratic, $P(I_s)$ degenerates to a Gaussian distribution. Since $\rho(\sqrt{x})$ has to be increasing and concave for $x\in (0, \infty)$ when $x$ follows the super-Gaussian distribution~\cite{palmer2010strong}, we can decouple $\rho()$ and $I_s$ using the concave conjugate of $\rho(\sqrt{F_x(I_s)})$ and  $\rho(\sqrt{F_y(I_s)})$ following the strategy of Babacan~\textit{et al.}~\cite{babacan2012bayesian}, and the upper bound of $\rho(F_x(I_s))$ and of $\rho(F_y(I_s))$ are represented as
\begin{equation}
\begin{aligned}
\rho(F_x(I_s)) \leq \frac{1}{2} \xi_x(F_x(I_s))^2 - \rho^\ast\left(\frac{1}{2} \xi_x\right),\\
\rho(F_y(I_s)) \leq \frac{1}{2} \xi_y(F_y(I_s))^2 - \rho^\ast\left(\frac{1}{2} \xi_y\right),
\label{eqn:upper_bound}
\end{aligned}
\end{equation}
where $\rho^\ast (\frac{1}{2} \xi_x)$ and $\rho^\ast (\frac{1}{2} \xi_y)$ denote the concave conjugates of $\rho(\sqrt{F_x(I_s)})$ and  $\rho(\sqrt{F_y(I_s)})$, respectively, and $\xi_x$ and $\xi_y$ are the variational parameters. We replace $\rho(F_x(I_s))$ and $\rho(F_y(I_s))$ in Eqn.~\ref{eqn:prior} with their upper bounds in $P(I_s)$
\begin{equation}
\begin{aligned}
P(I_s) \geq &W\exp\left( -\frac{\xi_x(F_x(I_s))^2 + \xi_y(F_y(I_s))^2}{4} \right) \\
&\cdot\exp \left(\frac{\rho^\ast (\frac{1}{2} \xi_x) + \rho^\ast (\frac{1}{2} \xi_y)}{2}\right).
\label{eqn:lower_bound}
\end{aligned}
\end{equation} 
Since the right-hand side of each inequality in Eqn.~\ref{eqn:upper_bound} is a convex quadratic function with a single global minimum, by calculating the derivative with respect to $F_x(I_s)$ and to $F_y(I_s)$, respectively, in Eqn.~\ref{eqn:upper_bound}, equality is attained when
\begin{gather}
\xi_x = \frac{\rho'(F_x(I_s))}{|F_x(I_s)|}, \xi_y =\frac{\rho'(F_y(I_s))}{|F_y(I_s)|},
\label{eqn:equality}
\end{gather}   
where $\rho'()$ is the derivative of $\rho()$. As shown in Eqn.~\ref{eqn:lower_bound}, irrespective of the form of $\rho()$, $P(I_s|\xi_x, \xi_y)$ becomes a trivial Gaussian distribution when equality is attained, which simplifies the derivation and the implementation because other penalty functions are discontinuous and the integral is too complicated to obtain (\textit{e.g.}, $|x|$, $\ln|x|$). Besides, a Gaussian distribution is usually utilized to approximate the real distribution in VB-based methods, and the multiplication of two Gaussian distributions is much easier to calculate.

\subsection{Variational Inference}
Due to the extra variational parameters $\xi_x$ and $\xi_y$, the problem can be reformulated as 
\begin{equation}
\begin{aligned}
&\underset {I_s, k, \xi_x, \xi_y} {\text{arg max}} P(I_s, k, \xi_x, \xi_y|I_b) \\
= &\underset {I_s, k, \xi_x, \xi_y} {\text{arg max}} \frac{P(I_b|I_s, k)P(I_s|\xi_x,\xi_y) P(\xi_x, \xi_y)P(k)}{P(I_b)}.
\label{eqn:map3}
\end{aligned}
\end{equation}
Directly calculating $P(I_s, k, \xi_x, \xi_y|I_b)$ is challenging because the true distribution of $I_b$ is difficult to obtain. The most common strategy is to use MAP, which estimates the posterior distribution by maximizing it as shown in Eqn.~\ref{eqn:map3}. However, as mentioned in Section~\ref{sec:introduction}, MAP with the sparse image prior favors a trivial solution. An alternative strategy is to use VB, which uses a trivial distribution $Q(I_s, k, \xi_x, \xi_y)$ (e.g., Gaussian) to approximate the posterior distribution $P(I_s, k, \xi_x, \xi_y|I_b)$ by minimizing the KL divergence between these two distributions, which can be written as
\begin{equation}
\begin{aligned}
&D_{KL} (Q(I_s, k, \xi_x, \xi_y)||P(I_s, k, \xi_x, \xi_y|I_b)) \\
=& \ln P(I_b) \\
&- \int Q(I_s, k, \xi_x, \xi_y)\ln \frac{P(I_s, k, \xi_x, \xi_y, I_b)}{Q(I_s, k, \xi_x, \xi_y)} dI_s dk d\xi_x d\xi_y\\
=& \ln P(I_b) - L(I_s, k, \xi_x, \xi_y, I_b),
\label{eqn:KL}
\end{aligned}
\end{equation}
where $D_{KL}$ represents the KL divergence, and  $L(I_s, k, \xi_x, \xi_y, I_b)$ is the variational lower bound. Since $\ln P(I_b) $ is constant and $D_{KL}$ is non-negative, minimizing $D_{KL}$ is equivalent to maximizing $L(I_s, k, \xi_x, \xi_y, I_b)$. By assuming that the $I_s$ and $k$ are independent, the variational lower bound can be rewritten as 
\begin{equation}
\begin{aligned}
&L(I_s, k, \xi_x, \xi_y, I_b)\\
=&\int Q(k)\ln \frac{P(k)}{Q(k)} dk - \int Q(I_s)\ln Q(I_s) dI_s \\
&+\int Q(I_s)Q(\xi_x,\xi_y)\ln P(I_s|\xi_x, \xi_y) dI_s d\xi_x d\xi_y\\
&+\int Q(\xi_x,\xi_y)\ln \frac{P(\xi_x,\xi_y) }{Q(\xi_x,\xi_y) } d\xi_x d\xi_y\\
&+ E_{Q(I_s, k)} \left[ \ln P(I_b|I_s, k) \right],
\label{eqn:LB}
\end{aligned}
\end{equation}
where $P(I_s|\xi_x, \xi_y)$ can be obtained from Eqn.~\ref{eqn:lower_bound}, $P(k)$ is set as the standard Gaussian distribution $\mathcal{N}(0,\, I)$. Based on the mean field theory~\cite{bishop2006pattern, babacan2012bayesian}, it is more convenient to simply assume that pixels on images and kernels are all independent. We can further rewrite Eqn.~\ref{eqn:LB} as
\begin{equation}
\begin{aligned}
&L(I_s, k, \xi_x, \xi_y, I_b)\\
=&\frac{1}{2} \sum_{i = 1}^{I}\sum_{j = 1}^{J} (2\ln S(k(i,j))- E^2(k(i,j)) - S^2(k(i,j)))\\
&+\frac{1}{2}\sum_{m = 1}^{M}\sum_{n = 1}^{N} 2\ln S(I_s(m, n))\\
&- \frac{1}{4} \sum_{m = 1}^{M}\sum_{n = 1}^{N} E((F_x(I_s) (m, n))^2) E(\xi_x(m,n))\\
&- \frac{1}{4} \sum_{m = 1}^{M}\sum_{n = 1}^{N} E((F_y(I_s) (m, n))^2) E(\xi_y(m,n))\\
&+ E_{Q(I_s, k)} \left[ \ln P(I_b|I_s, k) \right]\\
&+\int Q(\xi_x,\xi_y)\ln \frac{P(\xi_x,\xi_y)}{Q(\xi_x, \xi_y)} d\xi_xd\xi_y\\
&+\frac{1}{2}\int Q(\xi_x,\xi_y)(\rho^\ast (\frac{1}{2} \xi_x) + \rho^\ast (\frac{1}{2} \xi_y))d\xi_xd\xi_y\\
&+ Constant,
\label{eqn:LB_deep}
\end{aligned}
\end{equation}
where $S()$ and $E()$ denote the standard deviation and the expectation, respectively, of distribution $Q()$, $(i,j)$ is the pixel index of $k$, $(m,n)$ is the pixel index of $I_s$ and $\xi$. Since only the expectation of $\xi_x$ and $\xi_y$ are related to $I_s$, we do not need to consider their distributions so that the last three rows in Eqn.~\ref{eqn:LB_deep} can be ignored. Following Babacan~\textit{et al.}~\cite{babacan2012bayesian}, $E(\xi_x)$ and $E(\xi_y)$ can be simply calculated by
\begin{equation}
\begin{aligned}
E(\xi_x(m,n)) = \frac{\rho'(v_x(m,n))}{v_x(m,n)},\\
E(\xi_y(m,n)) = \frac{\rho'(v_y(m,n))}{v_y(m,n)},
\label{eqn:epsilon}
\end{aligned}
\end{equation}
\begin{equation}
\begin{aligned}
&v_x(m,n) = \sqrt{E((F_x(I_s) (m, n))^2)},\\
&v_y(m,n) = \sqrt{E((F_y(I_s) (m, n))^2)}.
\label{eqn:epsilon_2}
\end{aligned}
\end{equation}

For the sparse image prior, $F_x(I_s) (m, n)$ and $F_y(I_s) (m, n)$ can be reformulated as 
\begin{equation}
\begin{aligned}
&F_x(I_s) (m, n) = I_s(m,n)- I_s(m-1,n),\\
&F_y(I_s) (m, n) = I_s(m,n)- I_s(m,n-1),
\label{eqn:F_sparse}
\end{aligned}
\end{equation}
where $I_s(0,\cdot)$ and $I_s(\cdot, 0)$ denote paddings. 

Our VDIP can also be extended to the extreme channel prior. For the extreme channel prior, $F_x(I_s) (m, n)$ and $F_y(I_s) (m, n)$ can be reformulated as 
\begin{equation}
\begin{aligned}
&F_x(I_s) (m, n) = \underset{i\in \Omega(m, n)}{\text{min}} (\underset{c\in (r, g, b)}{\text{min}} (I^c_s (i))),\\
&F_y(I_s) (m, n) = 1 - \underset{i\in \Omega(m, n)}{\text{max}} (\underset{c\in (r, g, b)}{\text{max}} (I^c_s (i)))
\label{eqn:F_extreme}
\end{aligned}
\end{equation}
where $\Omega(m, n)$ denotes a local patch centered at $(m, n)$, and $I^c_s$ is a color channel of $I_s$. 

Further derivation of $E((F_x(I_s) (m, n))^2)$ and $E((F_y(I_s) (m, n))^2)$ are shown in the supplementary materials.

\subsection{Variational Deep Image Prior}
Conventional variational inference solves Eqn.~\ref{eqn:LB_deep} by calculating the closed-form expectation with respect to each variable over all the other variables to get the distribution~\cite{bishop2006pattern}, but it is challenging to apply this strategy to deep learning since the networks are highly non-convex. Hence, we use two networks to learn the distribution of the latent sharp image and the blur kernel, respectively, in an unsupervised manner. For simplification, we assume that the standard deviation of the blur kernel $S(k)$ is constant. We also assume that the additive noise is white Gaussian noise. Then, we only need to learn the expectation of the image $E(I_s)$, the expectation of the kernel $E(k)$, and the standard deviation of the image $S(I_s)$. 

\input{algorithm/alg_1}

We utilize an encoder-decoder as the image generator $G_I()$, a fully-connected network as the kernel generator $G_k()$, and random noises $Z_I$ and $Z_k$ as inputs. The image generator outputs both $E(I_s)$ and $S(I_s)$, and the kernel generator outputs $E(k)$. We can now approximate $E_{Q(I_s, k)} \left[ \ln P(I_b|I_s, k) \right]$ in Eqn.~\ref{eqn:LB} and~\ref{eqn:LB_deep} by Monte Carlo estimation using sampling~\cite{kingma2013auto}
\begin{equation}
\begin{aligned}
&E_{Q(I_s, k)} \left[ \ln P(I_b|I_s, k) \right] \approx \frac{1}{A}\sum_{a = 1}^{A} \frac{||I_b - \hat{k} \otimes \hat{I_s^a}||^2_2}{2\sigma^2},\\
&\hat{k} = E(k), \hat{I_s^a} = E(I_s) + \epsilon^a \odot S(I_s), \epsilon^a(m,n) \sim \mathcal{N} (0, \, I),
\end{aligned}
\label{eqn:expectation}
\end{equation}
where $A$ is the number of samples, $\sigma$ is the noise level, $\odot$ represents the element-wise multiplication, and $\epsilon^a(m,n)$ is a random scalar sampled from a standard Gaussian distribution for the pixel $(m,n)$. The more samplings, the more accurate distribution will be obtained. Using Monte Carlo estimation, the expectation term is now differentiable. Our final algorithm is shown in Alg.~\ref{alg:algorithm}.

The overview comparison of the DIP~\cite{ren2020neural} and our proposed method is shown in Fig.~\ref{fig:architecture}. We can see that the DIP only generates a single value $E(I_s)$ for each pixel instead of $E(I_s)$ and $S(I_s)$ in our VDIP, and the target is minimizing the mean square error $||I_b - E(k) \otimes E(I_s)||^2_2$. The target of the DIP only focuses on maximizing $P(I_b|I_s, k)$ in Eqn.~\ref{eqn:map_2}, so that $P(I_s|\theta_I)$ and $P(I_k|\theta_k)$ are not properly constrained. In contrast, in our proposed method, we apply a Gaussian prior and a sparse image prior to constrain $P(I_k|\theta_k)$ and $P(I_s|\theta_I)$, respectively, as shown in Eqn.~\ref{eqn:LB_deep}. Simply exploiting the additive priors for optimizing Eqn.~\ref{eqn:map_2} can lead to  suboptimal solutions of sparse MAP. Thus, we adpot the VB to avoid such a problem by introducing the standard deviation $S(I_s)$ to the optimization target. It is noteworthy that Eqn.~\ref{eqn:LB_deep} degenerates to the sparse MAP when we fix $S(I_s)$ as zero. It shows the limitation of optimizing the sparse MAP that its solution is difficult to achieve a large variational lower bound, because $\ln S(I_s)$ is negative infinity. The VB can nicely avoid it by considering non-zero $S(I_s)$. Besides, the values of $E(\xi)$ act as the penalty weights of gradients. In particular, small weights for large gradients and large weights for small gradients. Zero $S(I_s)$ may result in over-penalty in regions with small gradients. 

\input{figure/architecture}



%% file: algorithm/alg_1.tex
\begin{algorithm}[t]
	\caption{Blind Image Deconvolution Using Variational Deep Image Prior}
	\begin{algorithmic}
		\STATE \textbf{Input:} blurred image $I_b$, image generator $G_I()$, kernel generator $G_k()$ 
		\STATE \textbf{Output:} estimated sharp image $I_s^\ast$ and blur kernel $k^\ast$
		\STATE \textbf{Initialization:} fixed noise inputs $z_I$ and $z_k$, parameters of two generators $\theta_I^{(0)}$ and $\theta_k^{(0)}$ to be optimized
		\FOR{$t = 1, 2, \dots, T$}
		\STATE \parbox[t]{200pt}{1. generate $E(I_s)^{(t)}$, $S(I_s)^{(t)}$ by $G_I(z_I, \theta_I^{(t-1)})$ \\and $E(k)^{(t)}$ by  $G_k(z_k, \theta_k^{(t-1)})$\\}
		\STATE \parbox[t]{200pt}{2. calculate $E(\xi_x^{(t)})$ and $E(\xi_y^{(t)})$ using Eqn.~\ref{eqn:epsilon}\\}
		\STATE \parbox[t]{200pt}{3. sample $\hat{I_s}^{(t)}$ $A$ times and approximate \\ $E_{Q(I_s, k)} \left[ \ln P(I_b|I_s, k) \right]^{(t)}$ using Eqn.~\ref{eqn:expectation}\\}
		\STATE \parbox[t]{200pt}{4. calculate $L(I_s, k, \xi_x, \xi_y, I_b)^{(t)}$ using Eqn.~\ref{eqn:LB_deep}\\}
		\STATE 5. update $\theta_I^{(t-1)}$ and $\theta_k^{(t-1)}$ by maximizing \\$L(I_s, k, \xi_x, \xi_y, I_b)^{(t)}$
		\ENDFOR
		\STATE $[E(I_s)^{(T+1)}, S(I_s)^{(T+1)}]= G_I(z_I, \theta_I^{(T)})$
		\STATE $E(k)^{(T+1)} = G_k(z_k, \theta_k^{(T)})$
		\STATE $I_s^\ast = E(I_s)^{(T+1)}$, $k^\ast = E(k)^{(T+1)}$
	\end{algorithmic}
	\label{alg:algorithm}
\end{algorithm}

%% file: figure/architecture.tex
\begin{figure}[t]
  \centering
  \small
  \begin{tabular}{@{}c@{}}
    \includegraphics[width=\linewidth]{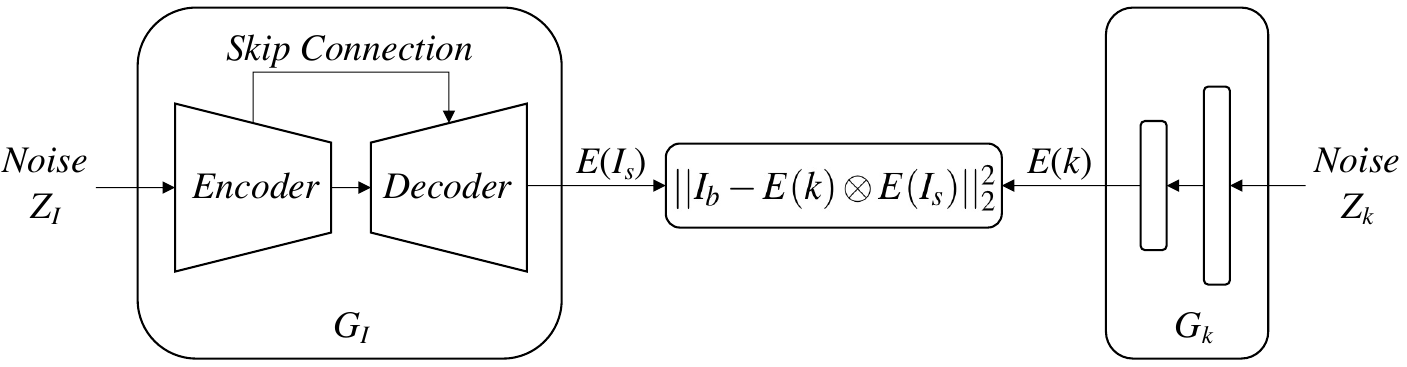} \\[\abovecaptionskip]
     (a) Overview of the DIP~\cite{ren2020neural}
  \end{tabular}

  \vspace{\floatsep}

  \begin{tabular}{@{}c@{}}
    \includegraphics[width=\linewidth]{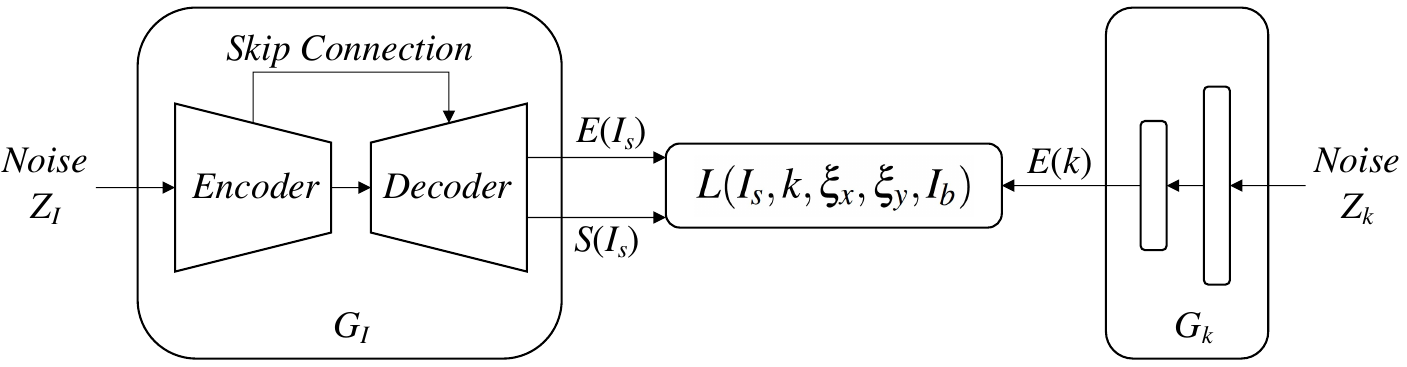} \\[\abovecaptionskip]
    (b) Overview of our proposed VDIP
  \end{tabular}

  \caption{Comparison of the DIP~\cite{ren2020neural} and our proposed VDIP. The number of decoder outputs are doubled and the loss function is replaced with the variational lower bound.}
  \label{fig:architecture}
\end{figure}

%% file: 5_experiments.tex
\section{Experiments}
\subsection{Implementation Details}
Our proposed method is implemented in PyTorch~\cite{paszke2019pytorch} and evaluated on a single RTX A6000 GPU with 48GB of memory. The learning rate of the image generator and of the kernel generator are set as $1\times10^{-2}$ and $1\times10^{-4}$, respectively, and the number of optimization steps $T$ is $5000$. In Eqn.~\ref{eqn:expectation}, the number of samples $A$ is set as 1. We use $\ln|x|$ as our penalty function $\rho(x)$. Note that the architectures of $G_I()$ and $G_k()$ are the same as those of DIP~\cite{ren2020neural} for fair comparison, except the output layers of $G_I()$ are doubled (half for $E(I_s)$ and half for $S(I_s)$). Different from the original DIP~\cite{ren2020neural} that adds additive random Gaussian noise to $Z_I$ and $Z_k$ to avoid the local minima, we do not add additive random noise to the inputs.

\input{table/comparison_1}

\input{table/comparison_3}

\input{table/comparison_2}

\subsection{Quantitative Comparison}

We first evaluate different versions of DIP including our VDIP for image deconvolution on the synthetic dataset from Lai~\textit{et al.}~\cite{lai2016comparative} and compare with several conventional methods including Cho and Lee~\cite{cho2009fast}, Levin~\textit{et al.}~\cite{levin2011efficient}, Krishnan~\textit{et al.}~\cite{krishnan2011blind}, Xu~\textit{et al.}~\cite{xu2013unnatural}, Perrone~\textit{et al.}~\cite{perrone2014total},
Michaeli and Irani~\cite{michaeli2014blind}, Pan~\textit{et al.}~\cite{pan2016blind}, Dong~\textit{et al.}~\cite{dong2017blind}, and Wen~\textit{et al.}~\cite{wen2020simple}, and several deep-learning-based methods including Tao~\textit{et al.}~\cite{tao2018scale}, Kupyn~\textit{et al.}~\cite{kupyn2019deblurgan}, Zamir~\textit{et al.}~\cite{zamir2021multi}, Huo~\textit{et al.}~\cite{huo2022blind}, Zamir~\textit{et al.}~\cite{zamir2022restormer} and Chen~\textit{et al.}~\cite{chen2022simple}. To be specific, these deep-learning-based methods are trained on external datasets~\cite{nah2017deep, nah2021ntire}. DIP-Extreme and DIP-Sparse represent the DIP~\cite{ren2020neural} with the extreme channel prior and the sparse image prior, respectively. Our VDIP-Std, VDIP-Extreme and VDIP-Sparse are the corresponding versions of DIP, DIP-Extreme and DIP-Sparse with non-zero $S(I_s)$.

The quantitative comparison is shown in Tab.~\ref{tab:comparison_1}. We can see that DIP-Sparse even perform worse than DIP, which is consistent with the suboptimal problem of sparse MAP. And non-zero $S(I_s)$ without additive priors can only slightly improve the performance. The combination of additive priors and non-zero $S(I_s)$ significantly increases the evaluation results, where the former helps to constrain the optimization and the latter avoids the local minimum resulting from the former. For gradient-based priors, a sparser constrain can lead to better performance comparing L0 norm~\cite{xu2013unnatural}, L1 norm~\cite{perrone2014total} and L2 norm~\cite{cho2009fast, levin2011efficient}, but the outliers on saturated images should be properly handled as in~\cite{dong2017blind}. Image-based priors~\cite{pan2016blind, wen2020simple} are more robust to outliers, and the comparison of DIP-Extreme and DIP-Sparse follows this observation. Additionally, our VDIP-sparse takes advantage of gradient-based priors without explicitly handling the outliers of saturated images and performs even better than VDIP-Extreme, which shows the effectiveness of utilizing variational Bayes.

To evaluate the estimated kernel, we calculate the average kernel recovery error~\cite{zhang2017global} and report the results in Tab.~\ref{tab:comparison_3}. Note that the compared deep-learning-based methods do not estimate the blur kernels. Although the evaluated kernel of Pan~\textit{et al.}~\cite{pan2016blind} is more accurate than VDIP-Extreme, our VDIP-Extreme performs better, which demonstrate that a proper deconvolution method is important even with accurate estimated blur kernels.

\input{figure/inference_time}

We also evaluate the above mentioned methods on the real blurred dataset from Lai~\textit{et al.}~\cite{lai2016comparative}. Since there is no ground truth sharp image, we utilize three no-reference image quality assessment metrics, in particular, Naturalness Image Quality Evaluator (NIQE)~\cite{mittal2012making}, Blind/Referenceless Image Spatial Quality Evaluator (BRISQUE)~\cite{mittal2011blind}, and Perception based Image Quality Evaluator (PIQE)~\cite{venkatanath2015blind} to quantitatively evaluate the results. As shown in Tab.~\ref{tab:comparison_2}, our method can generate images of the highest quality based on BRISQUE and PIQE among all compared methods. Similar to all of the compared conventional methods and DIP, our proposed method is also designed for spatially invariant (uniform) blur. However, it even performs better than deep-learning-based methods that are trained for spatially variant blur. We think this is because the compared deep-learning-based methods are all trained on synthetic datasets where the blurred images are generated by averaging consecutive frames from a high-frame-rate video. The performance of these methods are limited on the real data with more artifacts because of the domain-shift issue.

\subsection{Optimization Time}

To evaluate the relation between the optimization time and the size of images and kernels, we run the optimization with varying image size and fixed kernel size, and then run the optimization with varying kernel size and fixed image size. All of the experiments are run on a single RTX A6000 GPU with 48GB of memory. As shown in Fig.~\ref{fig:time}, the optimization time is proportional to the quadratic of image size and kernel size.

\subsection{Qualitative Comparison}

Some of the qualitative comparisons are shown in Fig.~\ref{fig:quantitative_1} and~\ref{fig:quantitative_2}. Our VDIP-Sparse can generate sharper results with less noise and artifacts than other methods including DIP. Specifically, Pan~\textit{et al.}~\cite{pan2016blind} are able to obtain correct blur kernels in some cases but the deconvolution results are over-smoothed. Dong~\textit{et al.}~\cite{dong2017blind}, Wen~\textit{et al.}~\cite{wen2020simple} and DIP~\cite{ren2020neural} are over-enhanced with many artifacts. Since the blur on the real images are spatially variant (non-uniform), obtaining perfect results with uniform deconvolution methods is difficult, if not impossible. But our method still performs better than Kupyn~\textit{et al.}~\cite{kupyn2019deblurgan} trained on non-uniform blurred datasets~\cite{nah2017deep}, showing the limited generalization ability of external training and the importance of image-specific information.

As outlined in Section~\ref{sec:introduction}, when a sparse image prior is employed in conjunction with Maximum a Posteriori (MAP) estimation, the resulting solution favors a trivial outcome, wherein the generated kernel is a delta kernel. Fig.~\ref{fig:quantitative_map} demonstrates the effectiveness of our improved approach utilizing Variational Bayes (VB) over the MAP method. It displays the trivial solution obtained by MAP, where the estimated kernels collapse to a single white dot (delta kernel). In contrast, VB successfully avoids such solutions, resulting in more accurate estimations.

\input{figure/qualitative_1}
\input{figure/qualitative_2}
\input{figure/qualitative_map}

\input{figure/failure}

\subsection{Failure Cases}

As shown in Fig.~\ref{fig:failure}, our VDIP does not perform well on small images with complex scenes, due to the lack of enough information to properly optimize the network.

%% file: table/comparison_1.tex
\begin{table*}[t]
\centering
\caption{Quantitative comparison (PSNR$\uparrow$/SSIM$\uparrow$) on the synthetic dataset from Lai~\textit{et al.}~\cite{lai2016comparative}.}
\label{tab:comparison_1}
\begin{tabular}{l|c|c|c|c|c|c}
\hline
\textbf{Method}  & \textbf{Manmade}     & \textbf{Natural}     & \textbf{People}        & \textbf{Saturated}      & \textbf{Text}   & \textbf{Average}     \\ \hline\hline
Cho~\textit{et al.}~\cite{cho2009fast} & 17.08/0.482  & 21.15/0.615  & 20.96/0.630  & 14.32/0.531  & 16.01/0.522  & 17.91/0.556 \\
Levin~\textit{et al.}~\cite{levin2011efficient}    & 15.12/0.284 & 18.76/0.419 & 19.55/0.528 & 13.98/0.487 & 14.44/0.372 & 16.37/0.418 \\ 
Krishnan~\textit{et al.}~\cite{krishnan2011blind} & 16.32/0.476 & 20.13/0.587 & 22.59/0.709 & 14.41/0.545 & 15.78/0.518 & 17.85/0.567 \\ 
Xu~\textit{et al.}~\cite{xu2013unnatural}       & 19.11/0.686 & 22.70/0.754 & 26.42/0.856 & 14.97/0.586 & 20.56/0.789 & 20.75/0.734 \\ 
Perrone~\textit{et al.}~\cite{perrone2014total}  & 18.66/0.676 & 22.78/0.786 & 24.79/0.828 & 14.46/0.531 & 18.35/0.673 & 19.81/0.699 \\ 
Michaeli~\textit{et al.}~\cite{michaeli2014blind} & 18.27/0.509& 21.93/0.614 & 25.74/0.791 & 14.46/0.539 & 16.59/0.503 & 19.40/0.591 \\
Pan~\textit{et al.}~\cite{pan2016blind}   & 20.00/0.714 & 24.47/0.801 & 26.70/0.811 & 17.46/0.680 & 21.13/0.762& 21.95/0.753\\ 
Dong~\textit{et al.}~\cite{dong2017blind}     & 18.88/0.567 & 23.42/0.702 & 25.53/0.769 & 16.72/0.611 & 20.05/0.682 & 20.92/0.666 \\ 
Tao~\textit{et al.}~\cite{tao2018scale}      & 17.11/0.381 & 20.18/0.492 & 22.12/0.651 & 15.41/0.545 & 15.76/0.469 & 18.12/0.508 \\ 
Kupyn~\textit{et al.}~\cite{kupyn2019deblurgan}    & 17.47/0.414 & 20.71/0.520 & 22.71/0.682 & 15.67/0.565 & 16.22/0.503 & 18.55/0.537 \\ 
Wen~\textit{et al.}~\cite{wen2020simple}      & 18.06/0.550 & 22.51/0.669 & 25.59/0.769& 17.79/0.672 & 17.85/0.598 & 20.36/0.652 \\ 
Zamir~\textit{et al.}~\cite{zamir2021multi}      & 17.12/0.392 & 20.30/0.506 & 21.50/0.631 & 15.49/0.547 & 14.75/0.415 & 17.83/0.498 \\
Huo~\textit{et al.}~\cite{huo2022blind}      & 17.11/0.380 & 20.27/0.495 & 21.69/0.636 & 15.45/0.545 & 15.84/0.478 & 18.07/0.507 \\
Zamir~\textit{et al.}~\cite{zamir2022restormer}      & 17.19/0.389 & 20.26/0.493 & 21.67/0.636 & 15.52/0.545 & 15.36/0.460 & 18.00/0.505 \\
Chen~\textit{et al.}~\cite{chen2022simple}      & 16.89/0.371 & 20.10/0.484 & 21.51/0.642 & 15.59/0.544 & 14.87/0.401 & 17.79/0.488 \\\hline\hline
Ren~\textit{et al.}~\cite{ren2020neural} (DIP) & 18.12/0.506  & 21.77/0.608  & 26.00/0.789  & 16.64/0.613  & 20.79/0.686 & 20.67/0.640 \\ 
DIP-Extreme  & 19.90/0.708  & 21.48/0.656  & 27.90/0.862  & 18.10/0.690  & 24.57/0.840 & 22.39/0.751 \\
DIP-Sparse  & 17.59/0.494  & 23.30/0.723  & 25.44/0.744  & 15.95/0.632  & 20.36/0.703 & 20.53/0.659 \\ \hline\hline
VDIP-Std  & 18.52/0.542  & 21.61/0.607  & 26.61/0.813  & 16.37/0.596  & 21.26/0.699 & 20.87/0.651 \\ 
VDIP-Extreme  & 20.50/0.768  & 25.36/0.882  & \textbf{30.83}/\textbf{0.938}  & 18.09/0.723  & 25.90/0.892 & 24.14/0.841 \\
VDIP-Sparse  & \textbf{22.86}/\textbf{0.868}  & \textbf{26.18}/\textbf{0.895}  & 30.76/0.927  & \textbf{18.55}/\textbf{0.727}  & \textbf{27.24}/\textbf{0.927} & \textbf{25.12}/\textbf{0.869} \\  \hline
\end{tabular}
\end{table*}

%% file: table/comparison_3.tex
\begin{table*}[t]
\centering
\caption{Average kernel recovery error on the synthetic dataset from Lai~\textit{et al.}~\cite{lai2016comparative}.}
\label{tab:comparison_3}
\begin{tabular}{l|c|c|c|c|c|c}
\hline
\textbf{Method}  & \textbf{Manmade}     & \textbf{Natural}     & \textbf{People}        & \textbf{Saturated}      & \textbf{Text}   & \textbf{Average}     \\ \hline\hline
Cho~\textit{et al.}~\cite{cho2009fast}   &0.00138 &0.00121 &0.00145 &0.00164 &0.00139 &0.00141  \\
Levin~\textit{et al.}~\cite{levin2011efficient}   &0.00099 &0.00107 &0.00117 &0.00124 &0.00117 &0.00113  \\
Krishnan~\textit{et al.}~\cite{krishnan2011blind}  &0.00125 &0.00114 &0.00128 &0.00134 &0.00118 &0.00124  \\
Xu~\textit{et al.}~\cite{xu2013unnatural}   &0.00114 &0.00084 &0.00073 &0.00144 &0.00074 &0.00098  \\
Perrone~\textit{et al.}~\cite{perrone2014total}   &0.00108 &0.00091 &0.00111 &0.00135 &0.00102 &0.00109  \\
Michaeli~\textit{et al.}~\cite{michaeli2014blind}  &0.00131 &0.00118 &0.00102 &0.00169 &0.00148 &0.00134  \\
Pan~\textit{et al.}~\cite{pan2016blind}   &0.00078 &\textbf{0.00060} &\textbf{0.00083} &\textbf{0.00099} &0.00071 &\textbf{0.00078}  \\
Dong~\textit{et al.}~\cite{dong2017blind}   &0.00097 &0.00078 &0.00096 &0.00111 &0.00082 &0.00093  \\
Wen~\textit{et al.}~\cite{wen2020simple}  &0.00113 &0.00092 &0.00089 &0.00074 &0.00098 &0.00093  \\ \hline\hline
Ren~\textit{et al.}~\cite{ren2020neural} (DIP)   &0.00168 &0.00168 &0.00164 &0.00172 &0.00144 &0.00163  \\
DIP-Extreme  &0.00117 &0.00122 &0.00084 &0.00153 &0.00086 &0.00113  \\
DIP-Sparse  &0.00159 &0.00148 &0.00136 &0.00142 &0.00135 &0.00144  \\ \hline\hline
VDIP-Std   &0.00163 &0.00167 &0.00157 &0.00171 &0.00140 &0.00160  \\
VDIP-Extreme  &0.00104 &0.00101 &0.00098 &0.00147 &0.00061 &0.00102  \\
VDIP-Sparse   &\textbf{0.00073} &0.00095 &0.00084 &0.00146 &\textbf{0.00060} &0.00092  \\ \hline
\end{tabular}
\end{table*}

%% file: table/comparison_2.tex
\begin{table}[t]
\centering
\caption{Quantitative comparison on the real blurred dataset from Lai~\textit{et al.}~\cite{lai2016comparative}.}
\label{tab:comparison_2}
\begin{tabular}{l|c|c|c}
\hline
\textbf{Method}   & \textbf{NIQE$\downarrow$}   & \textbf{BRISQUE$\downarrow$} & \textbf{PIQE$\downarrow$}    \\ \hline\hline
Cho~\textit{et al.}~\cite{cho2009fast}      & 4.0050 & 36.2829 & 48.6227 \\ 
Levin~\textit{et al.}~\cite{levin2011efficient}    & 3.6594 & 36.5006 & 46.7037 \\ 
Krishnan~\textit{et al.}~\cite{krishnan2011blind}  & 3.8696 & 37.9942 & 50.4024 \\ 
Xu~\textit{et al.}~\cite{xu2013unnatural}       & 3.9536 & 37.3240 & 49.5436 \\ 
Perrone~\textit{et al.}~\cite{perrone2014total}  & 4.0397 & 39.7997 & 51.7650 \\ 
Michaeli~\textit{et al.}~\cite{michaeli2014blind} & 3.5852 & 35.1205 & 46.7085 \\ 
Pan~\textit{et al.}~\cite{pan2016blind}  & 4.8790 & 36.3792 & 68.9470 \\ 
Dong~\textit{et al.}~\cite{dong2017blind}      & 4.7557 & 37.1199 & 64.1972 \\ 
Tao~\textit{et al.}~\cite{tao2018scale}      & 3.5612 & 40.1954 & 53.0908 \\ 
Kupyn~\textit{et al.}~\cite{kupyn2019deblurgan}    & \textbf{3.2937} & 35.8382 & 40.0545 \\ 
Wen~\textit{et al.}~\cite{wen2020simple}      & 4.9210 & 33.1731 & 58.3326 \\ 
Zamir~\textit{et al.}~\cite{zamir2021multi}      & 3.7926 & 42.4894 & 52.1181 \\
Huo~\textit{et al.}~\cite{huo2022blind}      & 3.5222 & 40.1037 & 47.0717 \\
Zamir~\textit{et al.}~\cite{zamir2022restormer}      & 3.7401 & 42.9266 & 50.6804 \\
Chen~\textit{et al.}~\cite{chen2022simple}      & 4.6754 & 46.3900 & 74.0267 \\\hline\hline
Ren~\textit{et al.}~\cite{ren2020neural} (DIP)    & 4.2460 & 38.5827 & 45.8822 \\ 
DIP-Extreme  & 4.7763 & 33.3678 & 36.6031 \\ 
DIP-Sparse   & 7.9063 & 41.9810 & 54.9295 \\ \hline\hline
VDIP-Std      & 4.1260 & 37.0199 & 42.3010 \\ 
VDIP-Extreme & 4.5072 & 34.4400 & 36.1535 \\ 
VDIP-Sparse  & 3.8882 & \textbf{32.4120} & \textbf{34.3614} \\ \hline
\end{tabular}
\end{table}

%% file: figure/inference_time.tex
\begin{figure}[t]
  \centering
    \includegraphics[width=\linewidth]{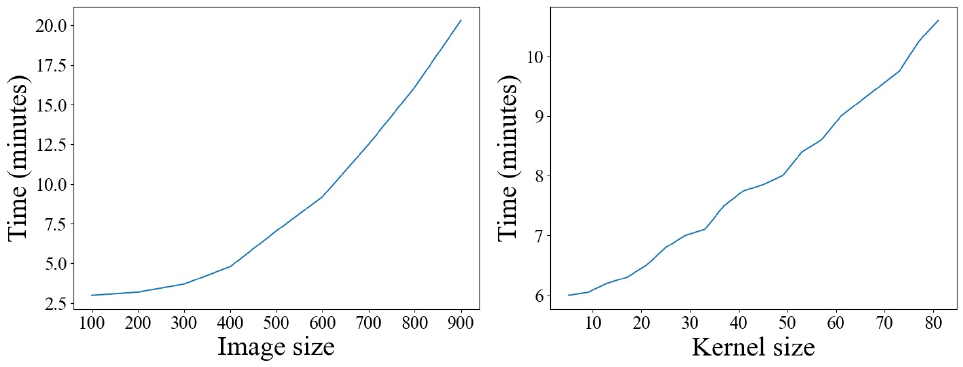} \\[\abovecaptionskip]
  \caption{The optimization time corresponding to the image size and kernel size. The kernel size is fixed as 31$\times$31 for evaluating the image size, and the image size is fixed as 500$\times$500 for evaluating the kernel size.}
  \label{fig:time}
\end{figure}

%% file: figure/qualitative_1.tex
\begin{figure*}[!htbp]
  \centering
    \includegraphics[width=\linewidth]{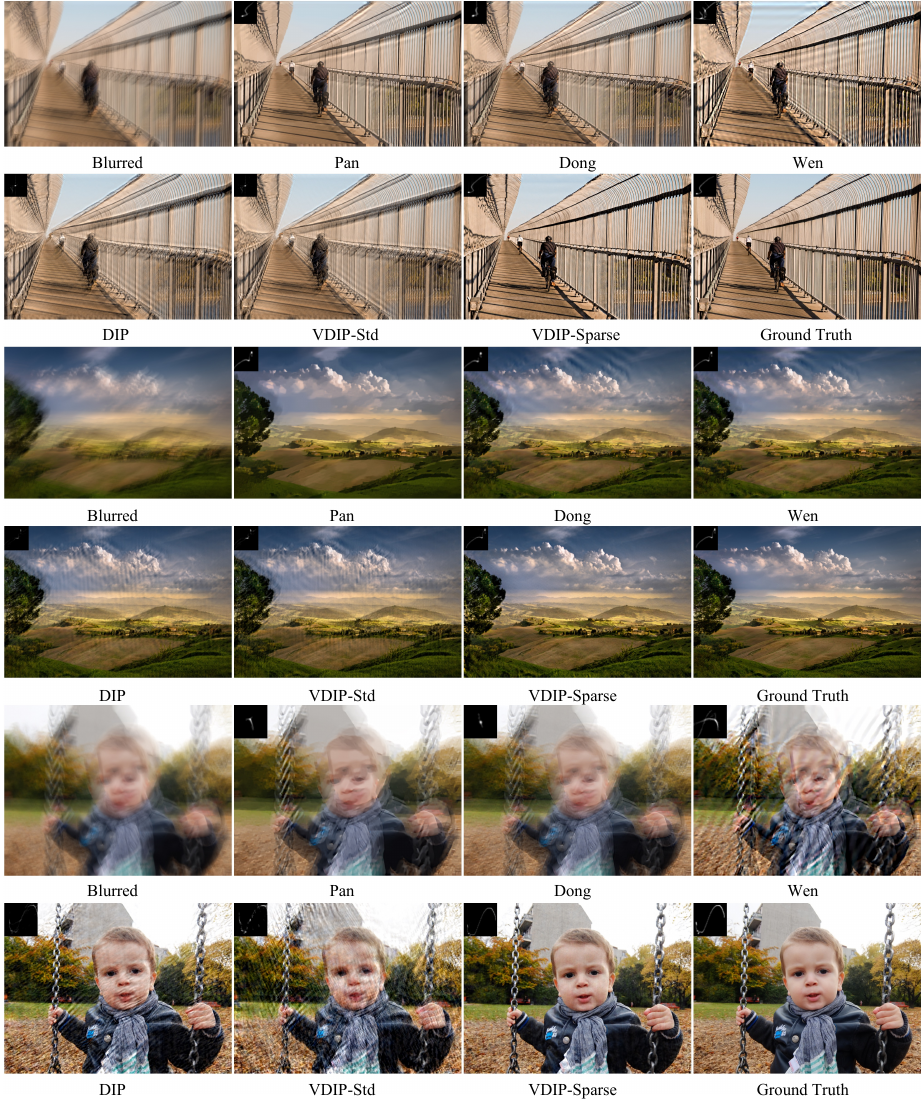} \\[\abovecaptionskip]
  \caption{Qualitative comparison on the synthetic dataset from Lai~\textit{et al.}~\cite{lai2016comparative}. The estimated blur kernels are pasted at the top-left corners of the corresponding deblurred results. }
  \label{fig:quantitative_1}
\end{figure*}

%% file: figure/qualitative_2.tex
\begin{figure*}[!htbp]
  \centering
    \includegraphics[width=0.9\linewidth]{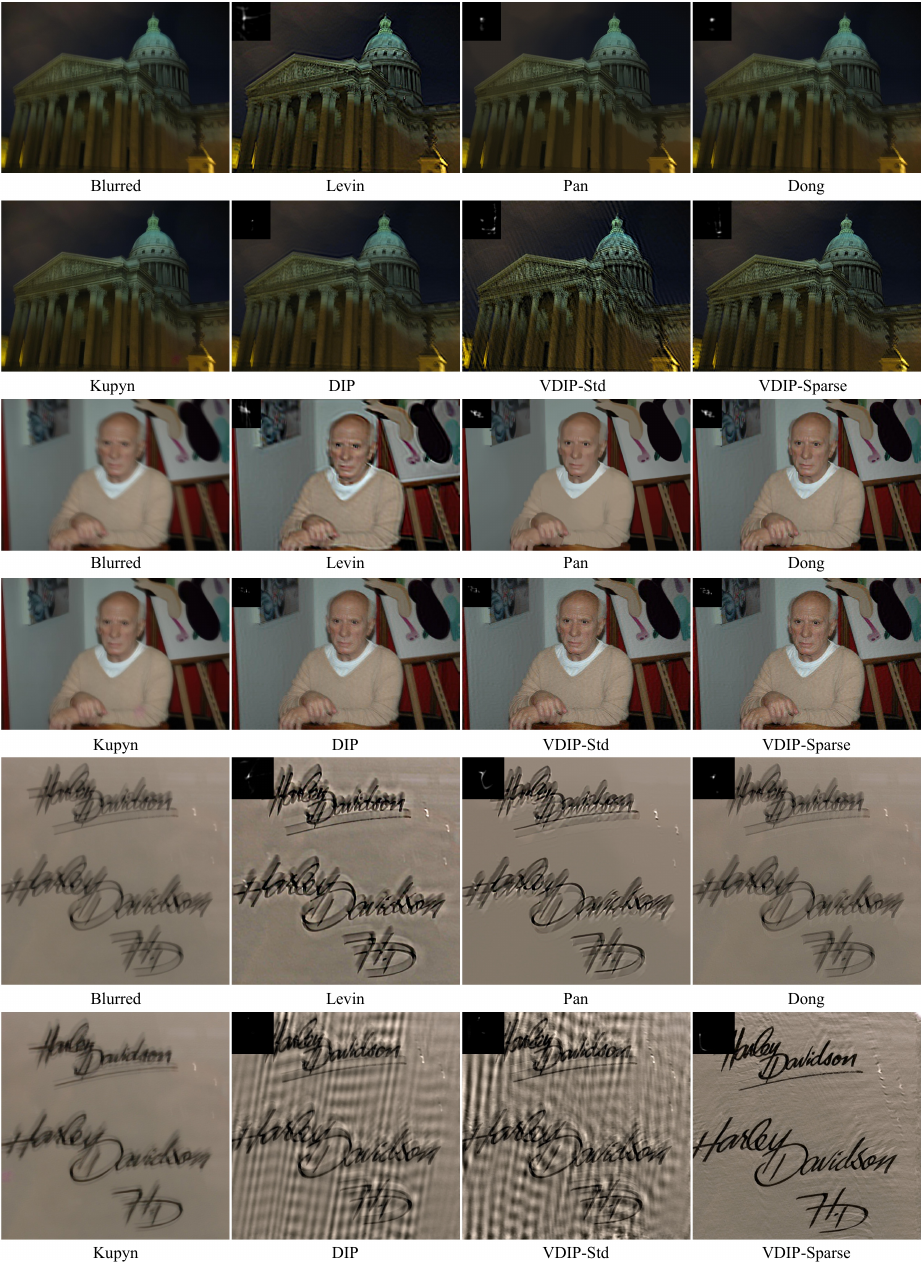} \\[\abovecaptionskip]
  \caption{Qualitative comparison on the real blurred dataset from Lai~\textit{et al.}~\cite{lai2016comparative}. The estimated blur kernels are pasted at the top-left corners of the corresponding deblurred results. }
  \label{fig:quantitative_2}
\end{figure*}

%% file: figure/qualitative_map.tex
\begin{figure*}[!htbp]
  \centering
    \includegraphics[width=0.99\linewidth]{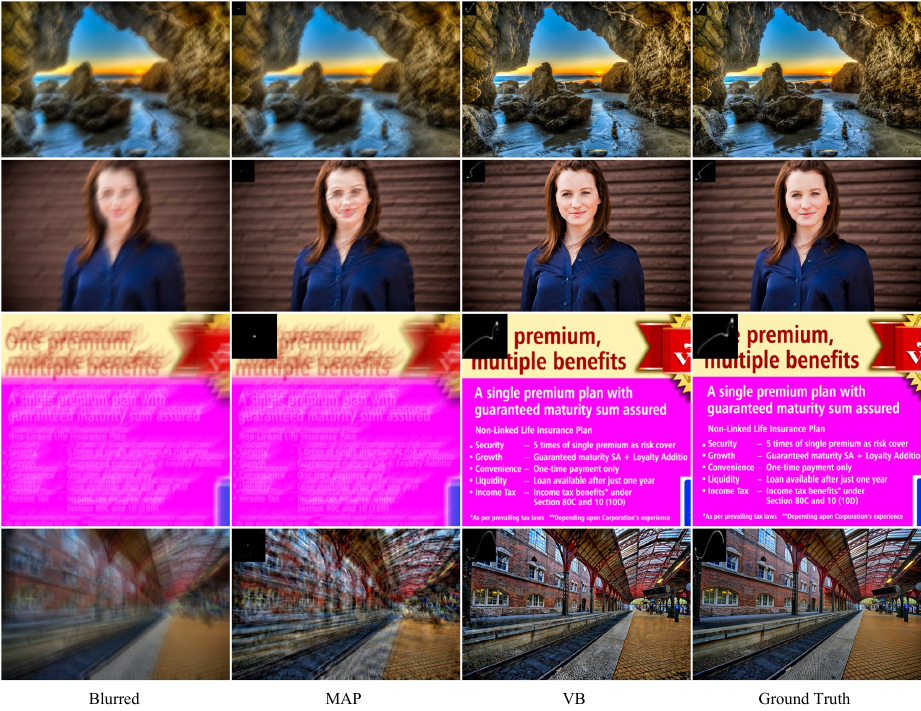} \\[\abovecaptionskip]
  \caption{Qualitative comparison of MAP (DIP) and VB (VDIP-Sparse). The estimated blur kernels are pasted at the top-left corners of the corresponding deblurred results where the estimated kernels of MAP are all delta kernels. }
  \label{fig:quantitative_map}
\end{figure*}

%% file: figure/failure.tex
\begin{figure*}[!htbp]
  \centering
    \includegraphics[width=0.99\linewidth]{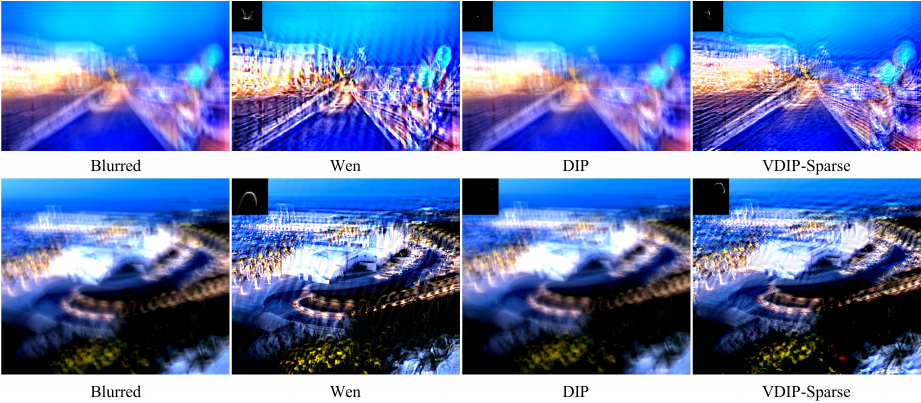} \\[\abovecaptionskip]
  \caption{Failure Cases.}
  \label{fig:failure}
\end{figure*}

%% file: 6_conclusion.tex
\section{Conclusion}
In this paper, we propose a new variational deep image prior (VDIP) for blind image deconvolution, which achieves a better performance than that of the DIP. One common issue of optimizing a model using a single image is high inference time compared with methods trained on external datasets, which makes it hard to adopt the method to large testing datasets. Our method is also limited when the single degraded image cannot provide enough information. In our future work, we plan to adopt meta-learning~\cite{soh2020meta} to train the networks on external datasets and fine-tune on each test image, which can take advantage of the information from other images and obtain a image-specific model with only several iterations. 

%% file: 7_acknowledgment.tex
\ifCLASSOPTIONcompsoc
  \section*{Acknowledgments}
\else
  \section*{Acknowledgment}
\fi

The authors would like to thank Steve Sutphen for his technical support and the Natural Sciences and Engineering Research Council of Canada, the Department of Computing Science, and the University of Alberta for funding. Masoumzadeh would like to thank Huawei for the Doctoral Scholarship.

%% file: 4_derivation.tex
\section{Detailed Derivation}\label{sec:derivation}
In this section, we show more detailed derivation of equations in Section 3. 
\subsection{Equation 9}

\begin{small}
\begin{align}
&D_{KL} (Q(I_s, k, \xi_x, \xi_y)||P(I_s, k, \xi_x, \xi_y|I_b)) \notag \\
=& \int Q(I_s, k, \xi_x, \xi_y)\ln\frac{Q(I_s, k, \xi_x, \xi_y)}{P(I_s, k, \xi_x, \xi_y|I_b)} dI_s dk d\xi_x d\xi_y\notag \\
=& \int Q(I_s, k, \xi_x, \xi_y)\ln\frac{Q(I_s, k, \xi_x, \xi_y) P(I_b)}{P(I_s, k, \xi_x, \xi_y, I_b)} dI_s dk d\xi_x d\xi_y\notag \\
=& \int Q(I_s, k, \xi_x, \xi_y)\ln P(I_b)dI_s dk d\xi_x d\xi_y - \int Q(I_s, k, \xi_x, \xi_y)\ln \frac{P(I_s, k, \xi_x, \xi_y, I_b)}{Q(I_s, k, \xi_x, \xi_y)} dI_s dk d\xi_x d\xi_y\notag \\
=& \ln P(I_b)\int Q(I_s, k, \xi_x, \xi_y)dI_s dk d\xi_x d\xi_y - \int Q(I_s, k, \xi_x, \xi_y)\ln \frac{P(I_s, k, \xi_x, \xi_y, I_b)}{Q(I_s, k, \xi_x, \xi_y)} dI_s dk d\xi_x d\xi_y\notag \\
=& \ln P(I_b)\notag - \int Q(I_s, k, \xi_x, \xi_y)\ln \frac{P(I_s, k, \xi_x, \xi_y, I_b)}{Q(I_s, k, \xi_x, \xi_y)} dI_s dk d\xi_x d\xi_y\notag \\
=& \ln P(I_b) - L(I_s, k, \xi_x, \xi_y, I_b). \tag{17}
\end{align}
\end{small}

\subsection{Equation 10}
\begin{small}
\begin{align}
&L(I_s, k, \xi_x, \xi_y, I_b)\notag \notag \\
=&\int Q(I_s, k, \xi_x, \xi_y)\ln \frac{P(I_s, k, \xi_x, \xi_y, I_b)}{Q(I_s, k, \xi_x, \xi_y)} dI_s dk d\xi_x d\xi_y\notag \\
=&\int Q(I_s, k, \xi_x, \xi_y)\ln \frac{P(I_s, k, \xi_x, \xi_y)}{Q(I_s, k, \xi_x, \xi_y)} dI_s dk d\xi_x d\xi_y + \int Q(I_s, k, \xi_x, \xi_y)\ln P(I_b|I_s, k, \xi_x, \xi_y) dI_s dk d\xi_x d\xi_y\notag\\
=&\int Q(I_s, k, \xi_x, \xi_y)\ln \frac{P(I_s, k, \xi_x, \xi_y)}{Q(I_s, k, \xi_x, \xi_y)} dI_s dk d\xi_x d\xi_y + E_{Q(I_s, k, \xi_x, \xi_y)} \left[ \ln P(I_b|I_s, k, \xi_x, \xi_y) \right]\notag \\
=&\int Q(I_s, k, \xi_x, \xi_y)\ln \frac{P(I_s, k, \xi_x, \xi_y)}{Q(I_s, k, \xi_x, \xi_y)} dI_s dk d\xi_x d\xi_y + E_{Q(I_s, k)} \left[ \ln P(I_b|I_s, k) \right]\notag \\
=&\int Q(I_s)Q(k)Q(\xi_x, \xi_y) \cdot \ln \frac{P(k)P(I_s|\xi_x, \xi_y)P(\xi_x, \xi_y)}{Q(I_s)Q(k)Q(\xi_x, \xi_y)} dI_s dk d\xi_x d\xi_y + E_{Q(I_s, k)} \left[ \ln P(I_b|I_s, k) \right]\notag \\
=&\int Q(I_s)Q(k)Q(\xi_x, \xi_y) \left(\ln \frac{P(k)}{Q(k)} - \ln Q(I_s) \right. + \left. \ln P(I_s|\xi_x, \xi_y) + \ln \frac{P(\xi_x,\xi_y) }{Q(\xi_x,\xi_y) }\right) dI_s dk d\xi_x d\xi_y + E_{Q(I_s, k)} \left[ \ln P(I_b|I_s, k) \right]\notag \\
=&\int Q(k)\ln \frac{P(k)}{Q(k)} dk - \int Q(I_s)\ln Q(I_s) dI_s + \int Q(I_s) Q(\xi_x,\xi_y) \ln P(I_s|\xi_x, \xi_y) dI_s d\xi_x d\xi_y +\int Q(\xi_x,\xi_y) \ln \frac{P(\xi_x,\xi_y) }{Q(\xi_x,\xi_y) } d\xi_x d\xi_y \notag \\
&+ E_{Q(I_s, k)} \left[ \ln P(I_b|I_s, k) \right].  \tag{18}
\end{align}
\end{small}
Since $\xi_x$ and $\xi_y$ are deterministic given $I_s$ following Eqn. 7, we can simply set $E_{Q(I_s, k, \xi_x, \xi_y)} \left[ \ln P(I_b|I_s, k, \xi_x, \xi_y) \right] = E_{Q(I_s, k)} \left[ \ln P(I_b|I_s, k) \right]$.

\subsection{Equation 11}

\begin{small}
\begin{align}
&\int Q(k)\ln \frac{P(k)}{Q(k)} dk \notag\\
=&\int \mathcal{N} (E(k), S^2(k)) \ln\frac{\mathcal{N} (0,\, I)}{\mathcal{N} (E(k), S^2(k))} dk\notag\\
=&\int \mathcal{N} (E(k), S^2(k)) \ln \mathcal{N} (0,\, I)dk - \int \mathcal{N} (E(k), S^2(k))\ln\mathcal{N} (E(k), S^2(k)) dk\notag\\
=&-\frac{1}{2} \sum_{i = 1}^{I}\sum_{j = 1}^{J}(\ln 2\pi + E^2(k(i,j)) + S^2(k(i,j))) + \frac{1}{2} \sum_{i = 1}^{I}\sum_{j = 1}^{J} (\ln 2\pi + 1 + 2\ln S(k(i,j)))\notag\\
=&\frac{1}{2} \sum_{i = 1}^{I}\sum_{j = 1}^{J} (1 + 2\ln S(k(i,j))- E^2(k(i,j)) - S^2(k(i,j))),  \tag{19}
\label{eqn:11_1}
\end{align}
\end{small}
$\mathcal{N}(E(), S^2())$ denotes the Gaussian distribution with mean $E()$ and variance $S^2()$, $S()$ and $E()$ denote the standard deviation and the expectation, respectively, $(i,j)$ is the pixel index of blur kernel.

\begin{small}
\begin{align}
&-\int Q(I_s)\ln Q(I_s) dI_s \notag\\
=& -\int \mathcal{N} (E(I_s), S^2(I_s))\ln\mathcal{N} (E(I_s), S^2(I_s)) dI_s\notag\\
=& \frac{1}{2} \sum_{m = 1}^{M}\sum_{n = 1}^{N} (\ln 2\pi + 1 + 2\ln S(I_s(m,n))), \tag{20}
\label{eqn:11_2}
\end{align}
\end{small}
$\mathcal{N}(E(), S^2())$ denotes the Gaussian distribution with mean $E()$ and variance $S^2()$, $S()$ and $E()$ denote the standard deviation and the expectation, respectively, $(m,n)$ is the pixel index of $I_s$ and $\xi$.

\begin{small}
\begin{align}
&\int Q(I_s) Q(\xi_x,\xi_y) \ln P(I_s|\xi_x, \xi_y) dI_s d\xi_x d\xi_y\notag\\
=&\int Q(I_s) Q(\xi_x,\xi_y)\left(\ln W  -\frac{\xi_x (F_x(I_s))^2 + \xi_y (F_y(I_s))^2}{4} \right. + \left. \frac{\rho^\ast (\frac{1}{2} \xi_x) + \rho^\ast (\frac{1}{2} \xi_y)}{2}\right) dI_s d\xi_x d\xi_y\notag\\
=&\ln W\int Q(I_s) Q(\xi_x,\xi_y) dI_s d\xi_x d\xi_y  -\int Q(I_s) Q(\xi_x,\xi_y)\left(\frac{\xi_x (F_x(I_s))^2 + \xi_y (F_y(I_s))^2}{4}\right) dI_s d\xi_x d\xi_y \notag\\
&+\int Q(I_s) Q(\xi_x,\xi_y)\left(\frac{\rho^\ast (\frac{1}{2} \xi_x) + \rho^\ast (\frac{1}{2} \xi_y)}{2}\right) dI_s d\xi_x d\xi_y\notag\\
=& -\int Q(I_s) Q(\xi_x,\xi_y)\left(\frac{\xi_x (F_x(I_s))^2 + \xi_y (F_y(I_s))^2}{4}\right) dI_s d\xi_x d\xi_y +\int Q(\xi_x,\xi_y)\left(\frac{\rho^\ast (\frac{1}{2} \xi_x) + \rho^\ast (\frac{1}{2} \xi_y)}{2}\right) d\xi_x d\xi_y + \ln W. \tag{21}
\label{eqn:11_3}
\end{align}
\end{small}

For the sparse image prior, $F_x(I_s)$ and $F_y(I_s)$ calculate the gradients of two directions as in Eqn. 14. 

Let us first look at the $F_x(I_s)$ related term in Eqn.~\ref{eqn:11_3}.

\begin{small}
\begin{align}
&-\int Q(I_s) Q(\xi_x,\xi_y)\frac{\xi_x (F_x(I_s))^2}{4} dI_s d\xi_x d\xi_y\notag\\
=& -\int Q(I_s) Q(\xi_x)\frac{\xi_x (F_x(I_s))^2}{4} dI_s d\xi_x\notag\\
=& -\int Q(I_s) Q(\xi_x)\frac{\sum_{m = 1}^{M}\sum_{n = 1}^{N}\xi_x(m,n) (F_x(I_s)(m,n))^2}{4} dI_s d\xi_x\notag\\
=& - \frac{1}{4} \sum_{m = 1}^{M}\sum_{n = 1}^{N} E((F_x(I_s) (m, n))^2) E(\xi_x(m,n)).\notag\\
=& - \frac{1}{4} \sum_{m = 1}^{M}\sum_{n = 1}^{N} E((I_s(m,n)-I_s(m-1,n))^2)E(\xi_x(m,n))\notag\\
=& -\frac{1}{4}\sum_{m = 1}^{M}\sum_{n = 1}^{N} E(\xi_x(m,n))[E^2(I_s(m,n)) + S^2(I_s(m,n)) - 2E(I_s(m,n))E(I_s(m-1,n)) + E^2(I_s(m-1,n)) + S^2(I_s(m-1,n))]\notag\\
=& - \frac{1}{4} \sum_{m = 1}^{M}\sum_{n = 1}^{N} [ (E(I_s(m,n))- E(I_s(m-1,n)))^2 + S^2(I_s(m,n)) + S^2(I_s(m-1,n))]E(\xi_x(m,n)). \tag{22}
\label{eqn:11_5}
\end{align}
\end{small}

The $F_y(I_s)$ related term can be derived in a similar way.
\begin{small}
\begin{align}
&-\int Q(I_s) Q(\xi_x,\xi_y)\frac{\xi_y (F_y(I_s))^2}{4} dI_s d\xi_x d\xi_y\notag\\
=&- \frac{1}{4} \sum_{m = 1}^{M}\sum_{n = 1}^{N} E((F_y(I_s) (m, n))^2) E(\xi_y(m,n))\notag\\
=&- \frac{1}{4} \sum_{m = 1}^{M}\sum_{n = 1}^{N} E((I_s(m,n)-I_s(m,n-1))^2)E(\xi_x(m,n))\notag\\
=& - \frac{1}{4} \sum_{m = 1}^{M}\sum_{n = 1}^{N} [ (E(I_s(m,n))- E(I_s(m,n-1)))^2 + S^2(I_s(m,n)) + S^2(I_s(m,n-1))]E(\xi_y(m,n)). \tag{23}
\label{eqn:11_6}
\end{align}
\end{small}

Combining Eqn. \ref{eqn:11_1} $\sim$ \ref{eqn:11_6}, we can get the variational lower bound as Eqn. 11.

Different from the sparse image prior which is differentiable and continuous, $F_x(I_s)$ and $F_y(I_s)$ (in Eqn. 15) are non-differentiable and discrete becasue of the $\text{max}()$ and the  $\text{min}()$. Thus, we cannot obtain closed-form expressions corresponding to $F_x(I_s)$ and $F_y(I_s)$ as in Eqn.~\ref{eqn:11_5} and Eqn~\ref{eqn:11_6}. To solve this problem, we  approximate  $E((F_x(I_s))^2)$ and $E((F_y(I_s))^2)$ by Monte Carlo estimation using sampling~\cite{kingma2013auto}.

Let us first look at the $F_x(I_s)$ related term in Eqn.~\ref{eqn:11_3}.

\begin{small}
\begin{align}
&-\int Q(I_s) Q(\xi_x,\xi_y)\frac{\xi_x (F_x(I_s))^2}{4} dI_s d\xi_x d\xi_y\notag\\
=& -\int Q(I_s) Q(\xi_x)\frac{\xi_x (F_x(I_s))^2}{4} dI_s d\xi_x\notag\\
=& -\int Q(I_s) Q(\xi_x)\frac{\sum_{m = 1}^{M}\sum_{n = 1}^{N}\xi_x(m,n) (F_x(I_s)(m,n))^2}{4} dI_s d\xi_x\notag\\
=& - \frac{1}{4} \sum_{m = 1}^{M}\sum_{n = 1}^{N} E((F_x(I_s) (m, n))^2) E(\xi_x(m,n)).\notag\\
=& - \frac{1}{4} \sum_{m = 1}^{M}\sum_{n = 1}^{N} E((\underset{i\in \Omega(m, n)}{\text{min}} (\underset{c\in (r, g, b)}{\text{min}} (I^c_s (i))))^2)E(\xi_x(m,n))\notag\\
=& - \frac{1}{4} \sum_{m = 1}^{M}\sum_{n = 1}^{N} E(\underset{i\in \Omega(m, n)}{\text{min}} (\underset{c\in (r, g, b)}{\text{min}} (I^c_s (i)^2)))E(\xi_x(m,n)). \tag{24}
\label{eqn:11_7}
\end{align}
\end{small}

Based on  Eqn.~\ref{eqn:11_7}, we need to approximate the expectation of $\underset{i\in \Omega(m, n)}{\text{min}} (\underset{c\in (r, g, b)}{\text{min}} (I^c_s (i)^2))$ by sampling  $I_s^2$. Thus, we need to calculate both  $E(I_s^2)$ and $S(I_s^2)$ as follows~\cite{normal}:
\begin{small}
\begin{align}
E(I_s^2) &= E(I_s)^2 + S(I_s)^2,\tag{25}\\
S(I_s^2) &= \sqrt{4E(I_s)^2 S(I_s)^2 + 2S(I_s)^4}. \tag{26}
\label{eqn:11_8}
\end{align}
\end{small}

Then the expectation can be reformulated as 
\begin{small}
\begin{align}
&E(\underset{i\in \Omega(m, n)}{\text{min}} (\underset{c\in (r, g, b)}{\text{min}} (I^c_s (i)^2))) \approx\frac{1}{A} \sum_{a = 1}^{A} \underset{i\in \Omega(m, n)}{\text{min}} (\underset{c\in (r, g, b)}{\text{min}} (\hat{I_s^2}^a(i)))) \notag\\
&\hat{I_s^2}^a = E(I_s^2) + \epsilon^a \odot S(I_s^2), \epsilon^a \sim \mathcal{N} (0, \, I), \tag{27}
\label{eqn:11_9}
\end{align}
\end{small}
where $A$ is the number of samples, $\sigma$ is the noise level, $\odot$ represents the element-wise multiplication, and $\epsilon^a$ is a random scalar sampled from a standard Gaussian distribution. 

The $F_y(I_s)$ related term in Eqn.~\ref{eqn:11_3} is slightly different.
\begin{small}
\begin{align}
&-\int Q(I_s) Q(\xi_x,\xi_y)\frac{\xi_y (F_y(I_s))^2}{4} dI_s d\xi_x d\xi_y\notag\\
=&- \frac{1}{4} \sum_{m = 1}^{M}\sum_{n = 1}^{N} E((F_y(I_s) (m, n))^2) E(\xi_y(m,n))\notag\\
=&- \frac{1}{4} \sum_{m = 1}^{M}\sum_{n = 1}^{N} E((1 - \underset{i\in \Omega(m, n)}{\text{max}} (\underset{c\in (r, g, b)}{\text{max}} (I^c_s (i))))^2) E(\xi_y(m,n))\notag\\
=&- \frac{1}{4} \sum_{m = 1}^{M}\sum_{n = 1}^{N} E((\underset{i\in \Omega(m, n)}{\text{min}} (\underset{c\in (r, g, b)}{\text{min}} (1 - I^c_s (i))))^2) E(\xi_y(m,n))\notag\\
=&- \frac{1}{4} \sum_{m = 1}^{M}\sum_{n = 1}^{N} E(\underset{i\in \Omega(m, n)}{\text{min}} (\underset{c\in (r, g, b)}{\text{min}} ((1 - I^c_s (i))^2))) E(\xi_y(m,n)). \tag{28}
\label{eqn:11_10}
\end{align}
\end{small}

Based on  Eqn.~\ref{eqn:11_10}, we need to approximate the expectation of $\underset{i\in \Omega(m, n)}{\text{min}} (\underset{c\in (r, g, b)}{\text{min}} ((1 - I^c_s (i))^2))$ by sampling $(1 - I_s)^2$. Similarly, we need to calculate both  $E((1 - I_s)^2)$ and $S((1 - I_s)^2)$ as follows:

\begin{small}
\begin{align}
E((1 - I_s)^2) &= (1 - E(I_s))^2 + S(I_s)^2,\tag{29}\\
S((1 - I_s)^2) &= \sqrt{4S(I_s)^2 + 4E(I_s)^2 S(I_s)^2 + 2S(I_s)^4}. \tag{30}
\label{eqn:11_11}
\end{align}
\end{small}

The form of expectation is the same as that shown in Eqn.~\ref{eqn:11_9} except that $I_s$ is replaced by $1 - I_s$.

%% file: main.bbl
\begin{thebibliography}{10}
\providecommand{\url}[1]{#1}
\csname url@samestyle\endcsname
\providecommand{\newblock}{\relax}
\providecommand{\bibinfo}[2]{#2}
\providecommand{\BIBentrySTDinterwordspacing}{\spaceskip=0pt\relax}
\providecommand{\BIBentryALTinterwordstretchfactor}{4}
\providecommand{\BIBentryALTinterwordspacing}{\spaceskip=\fontdimen2\font plus
\BIBentryALTinterwordstretchfactor\fontdimen3\font minus
  \fontdimen4\font\relax}
\providecommand{\BIBforeignlanguage}[2]{{%
\expandafter\ifx\csname l@#1\endcsname\relax
\typeout{** WARNING: IEEEtran.bst: No hyphenation pattern has been}%
\typeout{** loaded for the language `#1'. Using the pattern for}%
\typeout{** the default language instead.}%
\else
\language=\csname l@#1\endcsname
\fi
#2}}
\providecommand{\BIBdecl}{\relax}
\BIBdecl

\bibitem{joshi2008psf}
N.~Joshi, R.~Szeliski, and D.~J. Kriegman, ``Psf estimation using sharp edge
  prediction,'' in \emph{CVPR}, 2008.

\bibitem{shan2008high}
Q.~Shan, J.~Jia, and A.~Agarwala, ``High-quality motion deblurring from a
  single image,'' \emph{TOG}, vol.~27, no.~3, pp. 1--10, 2008.

\bibitem{chan1998total}
T.~F. Chan and C.-K. Wong, ``Total variation blind deconvolution,'' \emph{TIP},
  vol.~7, no.~3, pp. 370--375, 1998.

\bibitem{perrone2015clearer}
D.~Perrone and P.~Favaro, ``A clearer picture of total variation blind
  deconvolution,'' \emph{TPAMI}, vol.~38, no.~6, pp. 1041--1055, 2015.

\bibitem{vsroubek2020motion}
F.~{\v{S}}roubek and J.~Kotera, ``Motion blur prior,'' in \emph{ICIP}, 2020.

\bibitem{babacan2012bayesian}
S.~D. Babacan, R.~Molina, M.~N. Do, and A.~K. Katsaggelos, ``Bayesian blind
  deconvolution with general sparse image priors,'' in \emph{ECCV}, 2012.

\bibitem{krishnan2009fast}
D.~Krishnan and R.~Fergus, ``Fast image deconvolution using hyper-laplacian
  priors,'' \emph{NeurIPS}, 2009.

\bibitem{joshi2009image}
N.~Joshi, C.~L. Zitnick, R.~Szeliski, and D.~J. Kriegman, ``Image deblurring
  and denoising using color priors,'' in \emph{CVPR}, 2009.

\bibitem{xu2013unnatural}
L.~Xu, S.~Zheng, and J.~Jia, ``Unnatural l0 sparse representation for natural
  image deblurring,'' in \emph{CVPR}, 2013.

\bibitem{levin2009understanding}
A.~Levin, Y.~Weiss, F.~Durand, and W.~T. Freeman, ``Understanding and
  evaluating blind deconvolution algorithms,'' in \emph{CVPR}, 2009.

\bibitem{michaeli2014blind}
T.~Michaeli and M.~Irani, ``Blind deblurring using internal patch recurrence,''
  in \emph{ECCV}, 2014.

\bibitem{pan2016blind}
J.~Pan, D.~Sun, H.~Pfister, and M.-H. Yang, ``Blind image deblurring using dark
  channel prior,'' in \emph{CVPR}, 2016.

\bibitem{yan2017image}
Y.~Yan, W.~Ren, Y.~Guo, R.~Wang, and X.~Cao, ``Image deblurring via extreme
  channels prior,'' in \emph{CVPR}, 2017.

\bibitem{yang2021blind}
D.~Yang, X.-J. Wu, and H.~Yin, ``Blind image deblurring via enhanced sparse
  prior,'' \emph{Journal of Electronic Imaging}, vol.~30, no.~2, p. 023031,
  2021.

\bibitem{dong2017blind}
J.~Dong, J.~Pan, Z.~Su, and M.-H. Yang, ``Blind image deblurring with outlier
  handling,'' in \emph{ICCV}, 2017.

\bibitem{cho2009fast}
S.~Cho and S.~Lee, ``Fast motion deblurring,'' in \emph{SIGGRAPH Asia}, 2009.

\bibitem{chen2019blind}
L.~Chen, F.~Fang, T.~Wang, and G.~Zhang, ``Blind image deblurring with local
  maximum gradient prior,'' in \emph{CVPR}, 2019.

\bibitem{bai2019single}
Y.~Bai, H.~Jia, M.~Jiang, X.~Liu, X.~Xie, and W.~Gao, ``Single-image blind
  deblurring using multi-scale latent structure prior,'' \emph{CSVT}, vol.~30,
  no.~7, pp. 2033--2045, 2019.

\bibitem{fergus2006removing}
R.~Fergus, B.~Singh, A.~Hertzmann, S.~T. Roweis, and W.~T. Freeman, ``Removing
  camera shake from a single photograph,'' in \emph{SIGGRAPH}, 2006.

\bibitem{xu2010two}
L.~Xu and J.~Jia, ``Two-phase kernel estimation for robust motion deblurring,''
  in \emph{ECCV}, 2010.

\bibitem{nah2017deep}
S.~Nah, T.~Hyun~Kim, and K.~Mu~Lee, ``Deep multi-scale convolutional neural
  network for dynamic scene deblurring,'' in \emph{CVPR}, 2017.

\bibitem{nimisha2017blur}
T.~M. Nimisha, A.~Kumar~Singh, and A.~N. Rajagopalan, ``Blur-invariant deep
  learning for blind-deblurring,'' in \emph{ICCV}, 2017.

\bibitem{li2018learning}
L.~Li, J.~Pan, W.-S. Lai, C.~Gao, N.~Sang, and M.-H. Yang, ``Learning a
  discriminative prior for blind image deblurring,'' in \emph{CVPR}, 2018.

\bibitem{zhang2018dynamic}
J.~Zhang, J.~Pan, J.~Ren, Y.~Song, L.~Bao, R.~W. Lau, and M.-H. Yang, ``Dynamic
  scene deblurring using spatially variant recurrent neural networks,'' in
  \emph{CVPR}, 2018.

\bibitem{tao2018scale}
X.~Tao, H.~Gao, X.~Shen, J.~Wang, and J.~Jia, ``Scale-recurrent network for
  deep image deblurring,'' in \emph{CVPR}, 2018.

\bibitem{kupyn2018deblurgan}
O.~Kupyn, V.~Budzan, M.~Mykhailych, D.~Mishkin, and J.~Matas, ``Deblurgan:
  Blind motion deblurring using conditional adversarial networks,'' in
  \emph{CVPR}, 2018.

\bibitem{zhang2019gated}
X.~Zhang, H.~Dong, Z.~Hu, W.~S. Lai, F.~Wang, and M.~H. Yang, ``Gated fusion
  network for joint image deblurring and super-resolution,'' in \emph{BMVC},
  2019.

\bibitem{zhang2019deep}
H.~Zhang, Y.~Dai, H.~Li, and P.~Koniusz, ``Deep stacked hierarchical
  multi-patch network for image deblurring,'' in \emph{CVPR}, 2019.

\bibitem{lu2019unsupervised}
B.~Lu, J.-C. Chen, and R.~Chellappa, ``Unsupervised domain-specific deblurring
  via disentangled representations,'' in \emph{CVPR}, 2019.

\bibitem{kupyn2019deblurgan}
O.~Kupyn, T.~Martyniuk, J.~Wu, and Z.~Wang, ``Deblurgan-v2: Deblurring
  (orders-of-magnitude) faster and better,'' in \emph{ICCV}, 2019.

\bibitem{purohit2020region}
K.~Purohit and A.~Rajagopalan, ``Region-adaptive dense network for efficient
  motion deblurring,'' in \emph{AAAI}, 2020.

\bibitem{suin2020spatially}
M.~Suin, K.~Purohit, and A.~Rajagopalan, ``Spatially-attentive
  patch-hierarchical network for adaptive motion deblurring,'' in \emph{CVPR},
  2020.

\bibitem{zamir2021multi}
S.~W. Zamir, A.~Arora, S.~Khan, M.~Hayat, F.~S. Khan, M.-H. Yang, and L.~Shao,
  ``Multi-stage progressive image restoration,'' in \emph{CVPR}, 2021.

\bibitem{tran2021explore}
P.~Tran, A.~T. Tran, Q.~Phung, and M.~Hoai, ``Explore image deblurring via
  encoded blur kernel space,'' in \emph{CVPR}, 2021.

\bibitem{shocher2018zero}
A.~Shocher, N.~Cohen, and M.~Irani, ``“zero-shot” super-resolution using
  deep internal learning,'' in \emph{CVPR}, 2018.

\bibitem{ulyanov2018deep}
D.~Ulyanov, A.~Vedaldi, and V.~Lempitsky, ``Deep image prior,'' in \emph{CVPR},
  2018.

\bibitem{ren2020neural}
D.~Ren, K.~Zhang, Q.~Wang, Q.~Hu, and W.~Zuo, ``Neural blind deconvolution
  using deep priors,'' in \emph{CVPR}, 2020.

\bibitem{yang2019variational}
L.~Yang and H.~Ji, ``A variational em framework with adaptive edge selection
  for blind motion deblurring,'' in \emph{CVPR}, 2019.

\bibitem{kullback1997information}
S.~Kullback, \emph{Information theory and statistics}.\hskip 1em plus 0.5em
  minus 0.4em\relax Courier Corporation, 1997.

\bibitem{kingma2013auto}
D.~P. Kingma and M.~Welling, ``Auto-encoding variational bayes,'' in
  \emph{ICLR}, 2014.

\bibitem{krishnan2011blind}
D.~Krishnan, T.~Tay, and R.~Fergus, ``Blind deconvolution using a normalized
  sparsity measure,'' in \emph{CVPR}, 2011.

\bibitem{chen2020enhanced}
L.~Chen, F.~Fang, S.~Lei, F.~Li, and G.~Zhang, ``Enhanced sparse model for
  blind deblurring,'' in \emph{ECCV}, 2020.

\bibitem{lai2015blur}
W.-S. Lai, J.-J. Ding, Y.-Y. Lin, and Y.-Y. Chuang, ``Blur kernel estimation
  using normalized color-line prior,'' in \emph{CVPR}, 2015.

\bibitem{ren2016image}
W.~Ren, X.~Cao, J.~Pan, X.~Guo, W.~Zuo, and M.-H. Yang, ``Image deblurring via
  enhanced low-rank prior,'' \emph{TIP}, vol.~25, no.~7, pp. 3426--3437, 2016.

\bibitem{pan2019phase}
L.~Pan, R.~Hartley, M.~Liu, and Y.~Dai, ``Phase-only image based kernel
  estimation for single image blind deblurring,'' in \emph{CVPR}, 2019.

\bibitem{chakrabarti2016neural}
A.~Chakrabarti, ``A neural approach to blind motion deblurring,'' in
  \emph{ECCV}, 2016.

\bibitem{liu2016learning}
S.~Liu, J.~Pan, and M.-H. Yang, ``Learning recursive filters for low-level
  vision via a hybrid neural network,'' in \emph{ECCV}, 2016.

\bibitem{gong2017motion}
D.~Gong, J.~Yang, L.~Liu, Y.~Zhang, I.~Reid, C.~Shen, A.~Van Den~Hengel, and
  Q.~Shi, ``From motion blur to motion flow: A deep learning solution for
  removing heterogeneous motion blur,'' in \emph{CVPR}, 2017.

\bibitem{xu2017motion}
X.~Xu, J.~Pan, Y.-J. Zhang, and M.-H. Yang, ``Motion blur kernel estimation via
  deep learning,'' \emph{TIP}, vol.~27, no.~1, pp. 194--205, 2017.

\bibitem{liu2018recurrent}
J.~Liu, W.~Sun, and M.~Li, ``Recurrent conditional generative adversarial
  network for image deblurring,'' \emph{IEEE Access}, vol.~7, pp. 6186--6193,
  2018.

\bibitem{park2020multi}
D.~Park, D.~U. Kang, J.~Kim, and S.~Y. Chun, ``Multi-temporal recurrent neural
  networks for progressive non-uniform single image deblurring with incremental
  temporal training,'' in \emph{ECCV}, 2020.

\bibitem{asim2020blind}
M.~Asim, F.~Shamshad, and A.~Ahmed, ``Blind image deconvolution using deep
  generative priors,'' \emph{IEEE Transactions on Computational Imaging},
  vol.~6, pp. 1493--1506, 2020.

\bibitem{pan2020dgp_pami}
X.~Pan, X.~Zhan, B.~Dai, D.~Lin, C.~C. Loy, and P.~Luo, ``Exploiting deep
  generative prior for versatile image restoration and manipulation,''
  \emph{TPAMI}, pp. 1--1, 2021.

\bibitem{zhang2020plug}
K.~Zhang, Y.~Li, W.~Zuo, L.~Zhang, L.~Van~Gool, and R.~Timofte, ``Plug-and-play
  image restoration with deep denoiser prior,'' \emph{TPAMI}, 2021.

\bibitem{cheng2019bayesian}
Z.~Cheng, M.~Gadelha, S.~Maji, and D.~Sheldon, ``A bayesian perspective on the
  deep image prior,'' in \emph{CVPR}, 2019.

\bibitem{ho2020neural}
K.~Hoa, A.~Gilberta, H.~Jinb, and J.~Collomossea, ``Neural architecture search
  for deep image prior,'' \emph{Computers \& Graphics}, 2021.

\bibitem{gandelsman2019double}
Y.~Gandelsman, A.~Shocher, and M.~Irani, ``" double-dip": Unsupervised image
  decomposition via coupled deep-image-priors,'' in \emph{CVPR}, 2019.

\bibitem{abu2021image}
S.~Abu-Hussein, T.~Tirer, S.~Y. Chun, Y.~C. Eldar, and R.~Giryes, ``Image
  restoration by deep projected gsure,'' \emph{arXiv preprint
  arXiv:2102.02485}, 2021.

\bibitem{mataev2019deepred}
G.~Mataev, P.~Milanfar, and M.~Elad, ``Deepred: Deep image prior powered by
  red,'' in \emph{ICCV}, 2019.

\bibitem{vahdat2020NVAE}
A.~Vahdat and J.~Kautz, ``{NVAE}: A deep hierarchical variational
  autoencoder,'' in \emph{NeurIPS}, 2020.

\bibitem{palmer2010strong}
J.~A. Palmer, K.~Kreutz-Delgado, and S.~Makeig, ``Strong sub-and
  super-gaussianity,'' in \emph{International Conference on Latent Variable
  Analysis and Signal Separation}, 2010.

\bibitem{bishop2006pattern}
C.~M. Bishop, \emph{Pattern recognition and machine learning}.\hskip 1em plus
  0.5em minus 0.4em\relax springer, 2006.

\bibitem{paszke2019pytorch}
A.~Paszke, S.~Gross, F.~Massa, A.~Lerer, J.~Bradbury, G.~Chanan, T.~Killeen,
  Z.~Lin, N.~Gimelshein, L.~Antiga \emph{et~al.}, ``Pytorch: An imperative
  style, high-performance deep learning library,'' \emph{NeurIPS}, 2019.

\bibitem{lai2016comparative}
W.-S. Lai, J.-B. Huang, Z.~Hu, N.~Ahuja, and M.-H. Yang, ``A comparative study
  for single image blind deblurring,'' in \emph{CVPR}, 2016.

\bibitem{levin2011efficient}
A.~Levin, Y.~Weiss, F.~Durand, and W.~T. Freeman, ``Efficient marginal
  likelihood optimization in blind deconvolution,'' in \emph{CVPR}, 2011.

\bibitem{perrone2014total}
D.~Perrone and P.~Favaro, ``Total variation blind deconvolution: The devil is
  in the details,'' in \emph{CVPR}, 2014.

\bibitem{wen2020simple}
F.~Wen, R.~Ying, Y.~Liu, P.~Liu, and T.-K. Truong, ``A simple local minimal
  intensity prior and an improved algorithm for blind image deblurring,''
  \emph{CSVT}, 2020.

\bibitem{huo2022blind}
D.~Huo, A.~Masoumzadeh, and Y.-H. Yang, ``Blind non-uniform motion deblurring
  using atrous spatial pyramid deformable convolution and deblurring-reblurring
  consistency,'' in \emph{Proceedings of the IEEE/CVF Conference on Computer
  Vision and Pattern Recognition}, 2022, pp. 437--446.

\bibitem{zamir2022restormer}
S.~W. Zamir, A.~Arora, S.~Khan, M.~Hayat, F.~S. Khan, and M.-H. Yang,
  ``Restormer: Efficient transformer for high-resolution image restoration,''
  in \emph{Proceedings of the IEEE/CVF Conference on Computer Vision and
  Pattern Recognition}, 2022, pp. 5728--5739.

\bibitem{chen2022simple}
L.~Chen, X.~Chu, X.~Zhang, and J.~Sun, ``Simple baselines for image
  restoration,'' in \emph{Computer Vision--ECCV 2022: 17th European Conference,
  Tel Aviv, Israel, October 23--27, 2022, Proceedings, Part VII}.\hskip 1em
  plus 0.5em minus 0.4em\relax Springer, 2022, pp. 17--33.

\bibitem{nah2021ntire}
S.~Nah, S.~Son, S.~Lee, R.~Timofte, and K.~M. Lee, ``Ntire 2021 challenge on
  image deblurring,'' in \emph{Proceedings of the IEEE/CVF Conference on
  Computer Vision and Pattern Recognition}, 2021, pp. 149--165.

\bibitem{zhang2017global}
Y.~Zhang, Y.~Lau, H.-w. Kuo, S.~Cheung, A.~Pasupathy, and J.~Wright, ``On the
  global geometry of sphere-constrained sparse blind deconvolution,'' in
  \emph{CVPR}, 2017.

\bibitem{mittal2012making}
A.~Mittal, R.~Soundararajan, and A.~C. Bovik, ``Making a “completely blind”
  image quality analyzer,'' \emph{SPL}, vol.~20, no.~3, pp. 209--212, 2012.

\bibitem{mittal2011blind}
A.~Mittal, A.~K. Moorthy, and A.~C. Bovik, ``Blind/referenceless image spatial
  quality evaluator,'' in \emph{ASILOMAR}.\hskip 1em plus 0.5em minus
  0.4em\relax IEEE, 2011, pp. 723--727.

\bibitem{venkatanath2015blind}
N.~Venkatanath, D.~Praneeth, M.~C. Bh, S.~S. Channappayya, and S.~S. Medasani,
  ``Blind image quality evaluation using perception based features,'' in
  \emph{NCC}.\hskip 1em plus 0.5em minus 0.4em\relax IEEE, 2015, pp. 1--6.

\bibitem{soh2020meta}
J.~W. Soh, S.~Cho, and N.~I. Cho, ``Meta-transfer learning for zero-shot
  super-resolution,'' in \emph{CVPR}, 2020.

\bibitem{normal}
Wikipedia, ``Normal distribution,''
  \url{https://en.wikipedia.org/wiki/Normal_distribution}, 2022.

\end{thebibliography}
